\documentclass[useAMS,usenatbib]{mn2e}
\usepackage[totalwidth=515pt,totalheight=680pt,left=1.4cm,right=1.4cm]{geometry}
\usepackage{graphicx,amssymb}
\input psfig

\title[An Exoplanet's Response to Anisotropic Stellar Mass-Loss]
{An Exoplanet's Response to Anisotropic Stellar Mass-Loss During
Birth and Death}
\author[Veras, Hadjidemetriou and Tout]{Dimitri Veras$^{1}$\thanks{E-mail:
veras@ast.cam.ac.uk}, John D. Hadjidemetriou$^{2}$\thanks{Please see Acknowledgements} and Christopher A. Tout$^{1}$\\
$^{1}$Institute of Astronomy, University of Cambridge, Madingley Road, Cambridge CB3 0HA
\\
$^{2}$Department of Physics, Aristotle University of Thessaloniki, 54124 Thessaloniki, Greece
}

\begin{document}

\date{Accepted 2013 August 2.  Received 2013 July 23; in original form 2013 March 22}

\pagerange{\pageref{firstpage}--\pageref{lastpage}} \pubyear{XXXX} 

\maketitle

\label{firstpage}

\begin{abstract}
The birth and death of planets may be affected by mass outflows from their parent stars during the T-Tauri or post-main-sequence phases of stellar evolution.  These outflows are often modelled to be isotropic, but this assumption is not realistic for fast rotators, bipolar jets and supernovae.  Here we derive the general equations of motion for the time evolution of a single planet, brown dwarf, comet or asteroid perturbed by anisotropic mass loss in terms of a complete set of planetary orbital elements, the ejecta velocity, and the parent star's co-latitude and longitude.  We restrict our application of these equations to 1) rapidly rotating giant stars, and 2) arbitrarily-directed jet outflows.  We conclude that the isotropic mass-loss assumption can safely be used to model planetary motion during giant branch phases of stellar evolution within distances of hundreds of au.  In fact, latitudinal mass loss variations anisotropically affect planetary motion only if the mass loss is {\it asymmetric about the stellar equator}.  Also, we demonstrate how constant-velocity, asymmetric bipolar outflows in young systems incite orbital inclination changes.  Consequently, this phenomenon readily tilts exoplanetary orbits external to a nascent disc on the order of degrees.
\end{abstract}


\section{Introduction}

Planetary evolution and stellar evolution are intertwined.  A planet's birth disc may be periodically ingested by the parent star, triggering episodic mass outbursts that may strongly perturb the planet's motion.  Many Gyr later, during the giant branch phases of the star's post-main sequence (MS) evolution, the star sheds its mass, also affecting planetary motion.  One cannot develop a comprehensive theory for the life cycle of exoplanets without considering such mass-loss events.

Further, there is a strong observational motivation to explore such episodes of mass loss.  Mounting discoveries of planets orbiting post-MS stars are becoming too numerous to list individually in one paper; see Table 6 of \cite{getetal2012} for a reliable recent summary.  Explanations for their currently observed orbital parameters, as well as their past history and eventual fate, require accurate models of stellar mass loss in conjunction with orbital motion and potentially other effects such as tides. No planets have so far been confirmed orbiting white dwarfs.  This finding, as pointed out by \cite{musvil2012}, exemplifies the need to understand orbital evolution around giant stars so that future campaigns to search for planets around white dwarfs are motivated and well-directed.

The dearth of analyses of exoplanetary motion owing to mass loss has been partially alleviated in the last decade, primarily with post-MS studies.  Although the general two-body problem with mass loss has a long history (see \citealt*{rahetal2009} for a review), only recently have these analyses been directed towards understanding exoplanetary systems.  Of particular interest to exoplanetary scientists is the time evolution of orbital elements such as semimajor axis ($a$), eccentricity ($e$), inclination ($i$), longitude of ascending node ($\Omega$) and argument of pericentre ($\omega$).  This preference is partly reflected in the parameters which are reported in the major exoplanet databases (the Extrasolar Planets Encyclopedia at http://exoplanet.eu/, the Exoplanet Data Explorer at http://exoplanets.org/ and the NASA Exoplanet Archive at http://exoplanetarchive.ipac.caltech.edu/).  The earliest studies which presented expressions for this time evolution \citep{omarov1962,hadjidemetriou1963} helped lay the groundwork for future research conducted after the first hints of later-confirmed exoplanets \citep{cametal1988, latetal1989} and the first confirmed exoplanets \citep{wolfra1992}.  This research specifically includes effects which may cause an exoplanet to escape from a post-MS system \citep{veretal2011,vertou2012,adaetal2013} and generally encompasses studies investigating a planet engulfed by its expanding star \citep{villiv2007,villiv2009,noretal2010,villaver2011,musvil2012,pasetal2012,spiegel2012,norspi2012} as well as massive planet scattering in post-MS systems \citep{debsig2002,bonetal2011,bonetal2012,debetal2012,veretal2013,voyetal2013}.

However, in every one of these cases, mass loss is assumed to be isotropic.  Anisotropic mass-loss studies considering planetary motion are rare, and require care with respect to the reference frame used.  These studies include those by \cite{bouetal2012} and \cite{iorio2012}, both of whom consider anisotropic evaporation from planetary atmospheres as the driver for orbital evolution.  Investigators who instead consider asymmetric mass loss from the central star include \cite{paralc1998}, who quantify the escape of comets in different directions during post-MS evolution, and \cite{namouni2005}, \cite{namzho2006}, \cite{namguz2007}, and \cite{namouni2013}, who instead look at the effects of bipolar jets on planetary orbits during the early stages of the life of a planetary system.  

These early stages are also important to consider because the vast majority of known exoplanets are observed during MS evolution.  Therefore, their measured orbital parameters may have been strongly influenced by the parent star's behaviour while the protoplanetary birth disc was still present.  Systems with violent nebular histories featuring mass loss may be the source of some exoplanet eccentricities \citep{namouni2005}.  We will show how planetary inclinations may be similarly excited.  Although the vast majority of exoplanet inclinations are unknown, hints from highly inclined hot Jupiters \citep[e.g.][]{winetal2009,addetal2013,zhoche2013} as well as the Solar System illustrate that planets do not necessarily always form on planar orbits.


Here, we derive the complete equations of motion for the time evolution of planetary orbital elements accounting for latitude-, longitude- and time-dependent stellar mass loss.  Subsequently, by invoking symmetries in stellar mass loss models, we place strong constraints on the resulting orbital motion.   Although these equations have several applications, we choose two for this study.  The first is simply to quantify the goodness of the isotropic mass-loss approximation for planetary evolution in post-MS systems.   The second is to show how simply-approximated bipolar jets which are formed in the nascent stages of a planetary system can excite orbital inclinations.  We derive the equations of motion in Section 2 and apply them in Section 3 to post-MS evolution (Section 3.1) and pre-MS evolution (Section 3.2) before concluding in Section 4.

\section{Equations of Motion}

Our goal in this section is to express the time evolution of the orbital elements 
$(a,e,i,\Omega,\omega)$ in terms of one another and as a function of
position- and velocity-dependent mass ejection.  We begin by considering
the equations of motion in Cartesian coordinates.

\subsection{Inertial Frame}

Suppose a planet of constant mass $M_p$ is orbiting a solid body 
star of mass $M_s(t)$,
so that the total system mass is $M(t) = M_s(t)+M_p$. The stellar 
mass-loss flux, denoted by $J(\phi,\theta,t)>0$, is the mass-loss 
rate per steradian.  Here, $\phi$ is the stellar longitude
and $\theta$ is the stellar co-latitude.  Therefore,

\begin{equation}
\frac{dM(t)}{dt} = \frac{dM_s(t)}{dt} = -\frac{1}{4\pi} 
\int_{0}^{\pi}\int_{0}^{2\pi} J(\phi,\theta,t) \sin{\theta} d\phi d\theta
.
\end{equation}

The mass is lost at velocity $\vec u(\phi,\theta,t)$, 
where $\vec u$ is a vector along the radius of the star.  
Under the simplifying assumption that the direction of the rotational 
axis of the star is fixed with respect to the
orbit of the planet, then the equations of motion of the system are

\begin{eqnarray}
M_{p} \ddot{\vec{R}}_{p} 
&=&
\frac{G M_{p} M_{s}(t)}{\left| \vec{R}_{s} - \vec{R}_{p} \right|^3}
\left( \vec{R}_{s} - \vec{R}_{p}  \right)
,
\label{eqmotion1}
\\
M_{s}(t) \ddot{\vec{R}}_{s} 
&=&
\frac{G M_{s}(t) M_{p}}{\left| \vec{R}_{p} - \vec{R}_{s} \right|^3}
\left( \vec{R}_{p} - \vec{R}_{s}  \right)
+ \vec F
,
\label{eqmotion2}
\end{eqnarray}

\noindent{such that} $\vec{F} = \mathsf{Q} \vec{P}$, where 

\begin{equation}
\vec{P} =
\left(
\begin{array}{c}
\frac{1}{4\pi} \int_{0}^{\pi} \int_{0}^{2\pi} J(\phi,\theta,t) u(\phi,\theta,t) \cos{\phi} \sin{\theta} d\phi d\theta
\\
\frac{1}{4\pi} \int_{0}^{\pi} \int_{0}^{2\pi} J(\phi,\theta,t) u(\phi,\theta,t) \sin{\phi} \sin{\theta} d\phi d\theta
\\
\frac{1}{4\pi} \int_{0}^{\pi} \int_{0}^{2\pi} J(\phi,\theta,t) u(\phi,\theta,t) \cos{\theta}  d\phi d\theta
\end{array}
\right)
\end{equation}

\noindent{and} the matrix $\mathsf{Q}$ transforms from the spin from the star
to the orbital plane and has components

\begin{eqnarray}
Q_{11} &=& \cos{\Omega} \cos{\omega} - \sin{\Omega} \sin{\omega} \cos{i}
,
\\
Q_{21} &=& \sin{\Omega} \cos{\omega} + \cos{\Omega} \sin{\omega} \cos{i}
,
\\
Q_{31} &=& \sin{\omega} \sin{i}
,
\\
Q_{12} &=& -\cos{\Omega} \sin{\omega} - \sin{\Omega} \cos{\omega} \cos{i} 
,
\\
Q_{22} &=& -\sin{\Omega} \sin{\omega} + \cos{\Omega} \cos{\omega} \cos{i} 
,
\\
Q_{32} &=& \cos{\omega} \sin{i}
,
\\
Q_{13} &=& \sin{\Omega} \sin{i}
,
\\
Q_{23} &=& -\cos{\Omega} \sin{i}
,
\\
Q_{33} &=& \cos{i}
. 
\end{eqnarray}

\begin{figure}
\centerline{
\psfig{figure=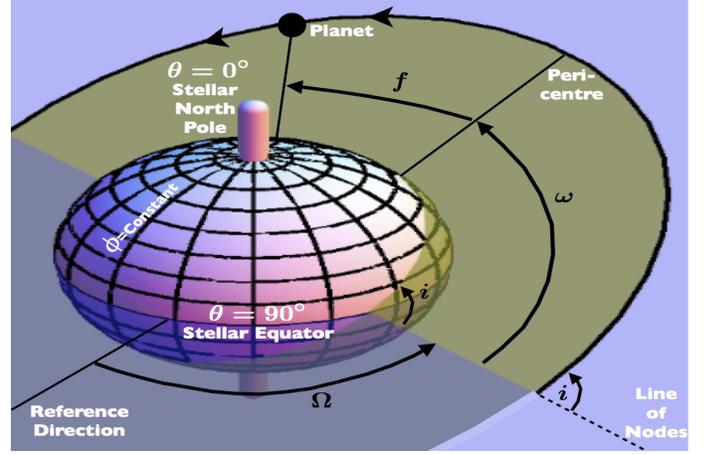,height=6.0cm,width=9.0cm} 
}
\caption{
Illustration displaying stellar co-latitude ($\theta$), stellar 
longitude ($\phi$), inclination ($i$), longitude of ascending 
node ($\Omega$), argument of pericentre ($\omega$) and 
true anomaly ($f$).}
\label{cartoon}
\end{figure}

\noindent{The} angles $\theta$, $\phi$, $i$, $\Omega$ and $\omega$ are illustrated
in Fig. \ref{cartoon}, along with the true anomaly $f$.

\subsection{Rotational Frame}

The equations of the relative motion can be obtained if 
we subtract equation (\ref{eqmotion2}) from equation (\ref{eqmotion1}). 
Defining the relative position of the planet with respect to 
the star by $\vec r$, where 

\begin{equation}
\vec r= \vec R_p -\vec R_s,
\end{equation}

\noindent{we} have

\begin {equation}
\ddot{\vec{r}} = -\frac{G M(t)}{r^3}\vec r - \frac{\vec F}{M_s(t)}.
\label{netpert}
\end{equation}

As a check, we can consider the isotropic mass-loss case.  
For isotropic mass-loss, $J$ and $\vec{u}$ are constant
for all $\phi$ and $\theta$ and are unchanging in time.  
Then, $\vec{P}=0$ and hence $\vec{F} = 0$.  
This result is expected because, 
in the isotropic case, the perturbation
to the motion due to mass-loss is implicit in $M_s(t)$.  
\cite{omarov1962}, \cite{hadjidemetriou1963} and 
\cite{deprit1983} found that this implicit perturbation to equal

\begin{equation}
\vec D = -\frac{1}{2} \frac{\dot{M}(t)}{M(t)} \vec{\dot{r}}
,
\label{pertP}
\end{equation}

\noindent{which} can be used to derive the equations of motion
for the planet in orbital elements in the isotropic case
\citep[e.g.][]{veretal2011}.

\subsection{Orbital Parameter Evolution}

Now, we wish to express the equations of motion in terms of the time
evolution of planetary orbital elements.  To do so, we use
the generalized perturbation equations of \cite{vereva2012}.  
The total perturbation to the two-body problem is then composed of 
both an implicit and explicit term, as in equations (\ref{netpert})-(\ref{pertP}),
and is equal to

\begin{equation}
\vec D - \frac{\vec F}{M_s}
.
\label{totpert}
\end{equation}

\noindent{We} can apply each of these terms
separately to the perturbation equations, and then add the result.
When doing so, we note importantly that although $\vec F$ is a function
of $\vec u$, $\vec F$ is not a function of the planet's position
nor its velocity.  We obtain

\begin{eqnarray}
\frac{da}{dt} &=& \left(\frac{da}{dt}\right)_{\rm iso} + \left(\frac{da}{dt}\right)_{\rm aniso}
,
\label{dadtorig}
\\
\frac{de}{dt} &=& \left(\frac{de}{dt}\right)_{\rm iso} + \left(\frac{de}{dt}\right)_{\rm aniso}
,
\\
\frac{di}{dt} &=& \left(\frac{di}{dt}\right)_{\rm iso} + \left(\frac{di}{dt}\right)_{\rm aniso}
,
\\
\frac{d\Omega}{dt} &=& \left(\frac{d\Omega}{dt}\right)_{\rm iso} + \left(\frac{d\Omega}{dt}\right)_{\rm aniso}
,
\\
\frac{d\omega}{dt} &=& \left(\frac{d\omega}{dt}\right)_{\rm iso} + \left(\frac{d\omega}{dt}\right)_{\rm aniso}
,
\\
\frac{df}{dt} &=& \left(\frac{df}{dt}\right)_{\rm unperturbed \ 2-body} 
                + \left(\frac{df}{dt}\right)_{\rm mass \ loss}
\nonumber
\\
&=& \frac{n \left(1 + e \cos{f}\right)^2}{\left(1 - e^2\right)^{3/2}}
                - \frac{d\omega}{dt} - \cos{i} \frac{d\Omega}{dt}
,
\label{dfdt}
\end{eqnarray}

\noindent{where} $n$ refers to the planet's mean motion.  The 
unperturbed 2-body term is the only term from 
equations (\ref{dadtorig})-(\ref{dfdt}) which survives when $J = 0$.

\subsubsection{General Equations}

By using the procedure of \cite{vereva2012} with the perturbation
given by equation (\ref{totpert}), 
we obtain the complete equations of motion for the planet.
The isotropic contributions are

\begin{eqnarray}
\left(\frac{da}{dt}\right)_{\rm iso} &=& -\frac{a \left(1 + e^2 + 2 e \cos{f}\right)}{1 - e^2}
\frac{1}{M} \frac{dM}{dt}
,
\\
\left(\frac{de}{dt}\right)_{\rm iso} &=& -\left(e + \cos{f} \right)
\frac{1}{M} \frac{dM}{dt}
,
\\
\left(\frac{di}{dt}\right)_{\rm iso} &=& 0
,
\\
\left(\frac{d\Omega}{dt}\right)_{\rm iso} &=& 0
,
\\
\left(\frac{d\omega}{dt}\right)_{\rm iso} &=& -\frac{\sin{f}}{e} \frac{1}{M} \frac{dM}{dt}
,
\\
\end{eqnarray}

\noindent{which} represents the standard set from \cite{omarov1962}, 
\cite{hadjidemetriou1963} and \cite{veretal2011}.  
These are the only surviving terms when $\vec{F} = 0$.  The non-isotropic
terms are

\[
\left(\frac{da}{dt}\right)_{\rm aniso} = -
\frac{2}{n M_s \sqrt{1-e^2}}
\bigg[
\nonumber
\]

\[
\ \ \ \ \ \ \ \ \ \ - \big\lbrace C_1 \cos{i} \sin{\Omega} + C_2 \cos{\Omega} \big\rbrace F_x
\nonumber
\]

\begin{equation}
\ \ \ \ \ \ \ \ \ \ + \big\lbrace C_1 \cos{i} \cos{\Omega} - C_2 \sin{\Omega} \big\rbrace F_y
+ \big\lbrace C_1 \sin{i} \big\rbrace F_z
\bigg]
,
\label{dadtgen}
\end{equation}

\[
\left(\frac{de}{dt}\right)_{\rm aniso} = -
\frac{\sqrt{1 - e^2}}{2 a n M_s \left(1 + e \cos{f} \right)}
\bigg[
\nonumber
\]

\[
\ \ \ \ \ \ \ \ \ \ - \big\lbrace C_6 \cos{i} \sin{\Omega} + C_5 \cos{\Omega} \big\rbrace F_x
\nonumber
\]

\begin{equation}
\ \ \ \ \ \ \ \ \ \ + \big\lbrace C_6 \cos{i} \cos{\Omega} - C_5 \sin{\Omega} \big\rbrace F_y
+ \big\lbrace C_6 \sin{i} \big\rbrace F_z
\bigg]
,
\end{equation}

\[
\left(\frac{di}{dt}\right)_{\rm aniso} = -\frac{\sqrt{1 - e^2}}{a n M_s \left(1 + e \cos{f} \right)}
\bigg[
\nonumber
\]

\[
\ \ \ \ \ \ \ \ \ \ +\big\lbrace\sin{i} \sin{\Omega} \cos{\left(f+ \omega\right)} \big\rbrace F_x
\nonumber
\]

\[
\ \ \ \ \ \ \ \ \ \ - \big\lbrace\sin{i} \cos{\Omega} \cos{\left(f+ \omega\right)} \big\rbrace F_y
\nonumber
\]

\begin{equation}
\ \ \ \ \ \ \ \ \ \
+ \big\lbrace\cos{i} \cos{\left(f+ \omega\right)} \big\rbrace F_z
\bigg]
,
\nonumber
\end{equation}

\[
\left(\frac{d\Omega}{dt}\right)_{\rm aniso} = -\frac{\sqrt{1 - e^2} \sin{\left(f+ \omega\right)} }{a n M_s \left(1 + e \cos{f} \right)}
\bigg[
\nonumber
\]

\begin{equation}
\ \ \ \ \ \ \ \ \ \ +\big\lbrace   \sin{\Omega} \big\rbrace F_x
- \big\lbrace   \cos{\Omega} \big\rbrace F_y
+ \big\lbrace   \cot{i}      \big\rbrace F_z
\bigg]
,
\nonumber
\end{equation}

\[
\left(\frac{d\omega}{dt}\right)_{\rm aniso} = -
\frac{\sqrt{1 - e^2}}{2 a e n M_s \left(1 + e \cos{f} \right)}
\bigg[
\nonumber
\]

\[
\ \ \ \ \ \ \ \ \ \ - \big\lbrace - C_8 \cos{i} \sin{\Omega} + C_7 \cos{\Omega} \big\rbrace F_x
\nonumber
\]

\[
\ \ \ \ \ \ \ \ \ \ - \big\lbrace C_8 \cos{i} \cos{\Omega} + C_7 \sin{\Omega} \big\rbrace F_y
\nonumber
\]

\begin{equation}
\ \ \ \ \ \ \ \ \ \ - \left\lbrace C_8 \sin{i} + e \sin{\left(f+ \omega\right) 
           \left[ \tan{\frac{i}{2}} + \cot{\frac{i}{2}} \right]  } \right\rbrace F_z
\bigg]
,
\label{domdtgen}
\end{equation}

\noindent{where} the auxiliary set of $C_i$ variables are:

\begin{eqnarray}
C_1 &\equiv& e \cos{\omega} + \cos{\left(f+\omega\right)}
,
\nonumber
\\
C_2 &\equiv& e \sin{\omega} + \sin{\left(f+\omega\right)}
,
\nonumber
\\
C_5 &\equiv& \left(3 + 4e \cos{f} + \cos{2f} \right) \sin{\omega}
        + 2 \left( e + \cos{f} \right) \cos{\omega} \sin{f}
,
\nonumber
\\
C_6 &\equiv& \left(3 + 4e \cos{f} + \cos{2f} \right) \cos{\omega}
        - 2 \left( e + \cos{f} \right) \sin{\omega} \sin{f}
,
\nonumber
\\
C_7 &\equiv& \left(3 + 2e \cos{f} - \cos{2f} \right) \cos{\omega}
        + \sin{\omega} \sin{2f}
,
\nonumber
\\
C_8 &\equiv& \left(3  - \cos{2f} \right) \sin{\omega}
        - 2 \left(e + \cos{f} \right) \cos{\omega} \sin{f} 
.
\nonumber
\end{eqnarray}

We wish to express equations (\ref{dadtgen})--(\ref{domdtgen})
in terms of properties of anisotropic stellar mass loss.  These are 
manifested in $\vec{P}$.  Inserting $\vec{F} = \mathsf{Q} \vec{P}$
into these equations yields remarkable simplifications

\begin{equation}
\left(\frac{da}{dt}\right)_{\rm aniso} = 
\frac{2}{n M_s \sqrt{1-e^2}}
\bigg[
\left( \sin{f} \right) P_x
- 
\left(e + \cos{f} \right) P_y
\bigg]
,
\label{dadtgenP}
\end{equation}

\[
\left(\frac{de}{dt}\right)_{\rm aniso} = 
\frac{\sqrt{1 - e^2}}{2 a n M_s \left(1 + e \cos{f} \right)}
\bigg[
\]

\begin{equation}
\ \ \ \ \ +2 \sin{f} \left(e + \cos{f} \right) P_x
-
\left(3 + 4 e \cos{f} + \cos{2f} \right) P_y
\bigg]
,
\label{dedtgenP}
\end{equation}

\begin{equation}
\left(\frac{di}{dt}\right)_{\rm aniso} = 
-\frac{\sqrt{1 - e^2} \cos{\left(f+\omega\right)}}{a n M_s \left(1 + e \cos{f} \right)}
P_z
,
\label{didtgenP}
\end{equation}

\begin{equation}
\left(\frac{d\Omega}{dt}\right)_{\rm aniso} = 
-\frac{\sqrt{1 - e^2} \sin{\left(f+ \omega\right)} \csc{i} }{a n M_s \left(1 + e \cos{f} \right)}
P_z
,
\label{dOmdtgenP}
\end{equation}

\[
\left(\frac{d\omega}{dt}\right)_{\rm aniso} = 
\frac{\sqrt{1 - e^2}}{2 a e n M_s \left(1 + e \cos{f} \right)}
\bigg[- \left(\sin{2f}\right) P_y
\]

\begin{equation}
\ \ \ \ \ +\left(3 + 2 e \cos{f} - \cos{2f} \right) P_x
+ 2 \cot{i} \sin{\left(f+\omega\right)} P_z
\bigg]
.
\label{domdtgenP}
\end{equation}

Equations (\ref{dadtgenP})--(\ref{domdtgenP}) are one of the 
main results of this study.  They show explicitly how the time evolution of
the orbital parameters relates to stellar mass loss at a given
latitude and longitude.  Although we have not found these relations
in existing literature, we can use previous studies to help
corroborate parts of our equations.  The most direct link 
is with \cite{omarov1962}.  The 
coefficients of the three $P_z$ terms in our
equations (\ref{didtgenP})--(\ref{domdtgenP})
are exactly equal to the coefficients of the $f_3$ terms in equation (13) of 
Omarov (\citeyear{omarov1962})
or equations (21)--(23) of the similar, 
English-language work by \cite{omarov1964}\footnote{The coefficients of the 
other components cannot be compared because of the different choices
of fiducial velocities chosen in equation (8) of \cite{omarov1962}
and equation (5) of \cite{vereva2012}.}.  Similarly, although
\cite{iorio2012} treats anisotropic mass loss from the planet
instead of the star and uses Gauss' planetary equations, which 
are derived with small perturbations, his 
coefficients for $di/dt$ and $d\Omega/dt$ in his equation (18) 
are equivalent to those in our equations 
(\ref{didtgenP})--(\ref{dOmdtgenP}).  \cite{bouetal2012}
also assume anisotropic mass loss from the planet,
apply a specific form for mass loss and do not utilize
the true anomaly.  Nevertheless, the dependencies in their 
equation (7) can be compared to those in our 
equations (\ref{dadtgenP})--(\ref{dedtgenP}).  

One can deduce several interesting properties from
our general equations of motion.

\begin{itemize}
\item{1) The non-isotropic terms become increasingly important the further
away the planet resides from the star because these terms contain an additional
factor of $\sqrt{a}$.}
\item{2) For a fixed total system mass, if the secondary is a brown 
dwarf or other type of star instead
of a planet, the effects of anisotropic mass loss are diminished.  Similarly
the importance of the anisotropic terms is maximized for test particle secondaries.} 
\item{3) The evolution of the semimajor axis, eccentricity and longitude of pericentre remain independent of $i$ and $\Omega$, as in the isotropic case.}
\item{4) A planet with an initially circular orbit does not retain that orbit.  Similarly, a planet's orbital inclination with respect to the stellar rotation axis is always changing. }
\end{itemize}

\subsubsection{Adiabatic Equations}

For many cases of interest, the planet's orbital period
is much smaller than the mass-loss 
timescale\footnote{In this approximation, the planetary orbit
behaves adiabatically in the sense that its eccentricity is
conserved; see \cite{debsig2002} and \cite{veretal2011}.}.  This comparison
remains true for a planet within a few hundred au of any
star that becomes a white dwarf \citep{veretal2011}.
In these cases, we can average over the true anomaly in 
equations (\ref{dadtgenP})-(\ref{domdtgenP})\footnote{
One should not average over equations (\ref{dadtgen})--(\ref{domdtgen})
first and then insert $\vec{F} = \mathsf{Q} \vec{P}$ into
the resulting equations.  Doing so yields unphysical dependencies
on $\vec{P}$.}.  We do so in the same manner as \cite{vereva2012},
where for an arbitrary variable $\beta$,

\begin{equation}
\frac{d\beta}{dt}_{\rm adiabatic} = \frac{n}{2\pi}
\int_{0}^{2\pi} \frac{d\beta}{dt}_{\rm non-adiabatic} \frac{dt}{df}df 
,
\end{equation}

\noindent{}and we assume 

\begin{equation}
\frac{dt}{df} = \frac{\left(1-e^2\right)^{3/2}}{n\left(1 + e\cos{f}\right)^2}
.
\end{equation}

\noindent{The} quality of this approximation for the isotropic case is
computed in equations (17--20) of \cite{veretal2011} and illustrated 
in fig. 1 of that paper.  They show that the adiabatic eccentricity 
variation from the initial value has an amplitude which is of the 
same order of the mass loss rate.  Averaging over the isotropic terms here yields

\begin{equation}
\left(\frac{da}{dt}\right)_{\rm iso}^{\rm adi} = 
-\frac{a}{M}
\frac{dM}{dt}
,
\end{equation}

\begin{equation}
\left(\frac{de}{dt}\right)_{\rm iso}^{\rm adi} = 
\left(\frac{di}{dt}\right)_{\rm iso}^{\rm adi} = 
\left(\frac{d\Omega}{dt}\right)_{\rm iso}^{\rm adi} = 
\left(\frac{d\omega}{dt}\right)_{\rm iso}^{\rm adi} = 
0
.
\end{equation}

Conversely, averaging over the non-isotropic terms
causes only the $da/dt$ term to vanish so that

\begin{equation}
\left(\frac{da}{dt}\right)_{\rm aniso}^{\rm adi} = 0
,
\label{dadtP}
\end{equation}

\begin{equation}
\left(\frac{de}{dt}\right)_{\rm aniso}^{\rm adi} = 
- \frac{3 \sqrt{1 - e^2}}{2 a n M_s} P_{y}
,
\label{dedtP}
\end{equation}

\begin{equation}
\left(\frac{di}{dt}\right)_{\rm aniso}^{\rm adi} = 
\frac{3e \cos{\omega}}{2 a n M_s\sqrt{1 - e^2}} P_{z}
,
\end{equation}

\begin{equation}
\left(\frac{d\Omega}{dt}\right)_{\rm aniso}^{\rm adi} = 
-\frac{3e \sin{\omega} \csc{i}}{2 a n M_s\sqrt{1 - e^2}} P_{z}
,
\end{equation}

\begin{eqnarray}
\left(\frac{d\omega}{dt}\right)_{\rm aniso}^{\rm adi} 
&=& 
\frac{3}{2 a e n M_s} 
\nonumber
\\
&\times&
\big(
\sqrt{1 - e^2} P_{x}
-
\frac{e^2}{\sqrt{1 - e^2}}
\sin{\omega} \cot{i}
P_{z}
\big)
.
\label{domdtP}
\end{eqnarray}

\noindent{Equations} (\ref{dadtP})--(\ref{domdtP}) are another
main result of this paper and allow us to deduce
planetary orbital properties directly from the character
of anisotropic stellar mass loss in $\vec{P}$
for most planets within a few hundred au of their parent
star.  For planets or comets which are further away,
we must use equations (\ref{dfdt}) and 
(\ref{dadtgenP})--(\ref{domdtgenP}).  Properties of the
adiabatic equations include the following.

\begin{itemize}
\item{1) Instability may occur, unlike the adiabatic isotropic case.  Regardless of 
the planet's semimajor axis, its eccentricity may approach unity and hence collide
with the star.}
\item{2) Anisotropic terms are retained after averaging in each equation except for
$da/dt$, and the time evolution of $e$, $I$, $\Omega$ and $\omega$ are each proportional
to $\sqrt{a}$.} 
\item{3) Unlike the isotropic case, initially circular or coplanar orbits are not
guaranteed to remain so.}
\end{itemize}

\section{Applications}

Here we present examples of how the above equations
might be applied to studies focused on a particular
exosystem.  

\subsection{Post-MS Rotation}

We first consider planetary dynamical evolution in post-MS systems,
where stellar rotation is linked to mass loss.  We
discuss the dependence on co-latitude and then longitude
before moving on to specific examples.

\subsubsection{Latitudinal Dependence}

Several authors have investigated physically sensible scaling laws 
for mass loss and velocity ejecta as a function of $\theta$.
In a seminal paper describing stellar winds, \cite{casetal1975} establish
a scaling law for mass loss (their equation 46) in terms of intrinsic stellar
properties. \cite{owoetal1998} re-expressed this law in a form more
useful for our purposes,

\begin{equation}
J
=
J\left(\theta=0,t\right)
\left[
\frac{H(\theta)}{H(\theta = 0)}
\right]^{\frac{1}{\alpha}}
\left[
1 - \frac{V_{\rm rot}^2 p_s}{G M_s(t)} \sin^2{\theta}
\right]^{1 - \frac{1}{\alpha}}
,
\label{Jeq}
\end{equation}

\noindent{where} $p_s$ is the radius of the star,
$V_{\rm rot}$ is the rotational velocity of the star at the equator,
$H$ is the radiative flux and $\alpha$ is an empirically-determined 
constant.  \cite{puletal2008} provide recent estimates of 
$0.6 \lesssim \alpha \lesssim 0.7$
for O-type stars and $\alpha \approx 0.45$ for A supergiants.
These estimates broadly agree with more recent adoptions of $\alpha \approx 0.60$ 
\citep{lovekin2011} and $\alpha = 0.43$ \citep{geoetal2011}.
However, we can simplify the expression for $J\left(\phi,\theta,t\right)$ and thereby eliminate the dependence on $\alpha$ by considering the gravity
darkening law of \cite{vonzeipel1924} so that

\begin{equation}
\frac{H(\theta)}{H(\theta = 0)}
=
1 - \frac{V_{\rm rot}^2 p_s}
{G M_s(t)}
\sin^2{\theta}
.
\label{Heq}
\end{equation}

\cite{owoetal1998} also provide an estimate for latitudinally-varying
ejecta speed

\begin{equation}
u(\phi,\theta,t) = \sqrt{ 
\left( \frac{2 G M_s(t)}{p_s} \right)
\left(
1 - 
\frac{V_{\rm rot}^2 p_s}
{G M_s(t)}
\sin^2{\theta}
\right)
    }
.
\label{ueq}
\end{equation}

Equations (\ref{Jeq})--(\ref{ueq}) have no longitudinal dependence.  Consequently,
the double integrals in the expression for $\vec{F}$ become

\begin{eqnarray}
P_x
&=&
\frac{1}{4\pi}
\int_{0}^{2\pi} \cos{\phi} d\phi
\int_{0}^{\pi}  J(\theta,t) u(\theta,t)  \sin{\theta}  d\theta
= 
0
,
\label{x0}
\\
P_y
&=&
\frac{1}{4\pi}
\int_{0}^{2\pi} \sin{\phi} d\phi
\int_{0}^{\pi} J(\theta,t) u(\theta,t) \sin{\theta} d\theta
=
0
,
\label{y0}
\\
P_z
&=&
\frac{1}{4\pi}  
\int_{0}^{2\pi} d\phi
\int_{0}^{\pi} J(\theta,t) u(\theta,t) \cos{\theta}  d\theta
\nonumber
=\frac{1}{2} J\left(\theta=0,t\right)
\nonumber
\end{eqnarray}

\begin{equation}
\ \ \ \times
\sqrt{ 
\frac{2 G M_s(t)}{p_s} 
}
\int_{0}^{\pi}
\left[
1 - \frac{V_{\rm rot}^2 p_s}
{G M_s(t)}
\sin^2{\theta}
\right]^{\frac{3}{2}}
\cos{\theta} d\theta
=0
.
\label{z0}
\end{equation}

\noindent{The} last integral shows that despite the latitudinal dependencies, mass loss which is
symmetric about the stellar equator has no net effect on a planet's motion in addition to the usual
effect from isotropic mass loss.  This result helps verify the robustness of the isotropic
mass loss approximation in post-MS studies.

Nevertheless, let us consider the possibility that mass-loss is not symmetric about the equator.
Here we assume that the mass loss from the northern hemisphere of the star is $k>1$ times
the amount of mass lost from the southern hemisphere and adopt the same latitudinal dependence
as in equation~(\ref{Jeq}) and velocity dependence as in equation (\ref{ueq}).  Also, let us 
describe the square of the ratio of the rotational velocity to the circular velocity at the stellar 
surface by $T(t) \equiv V_{\rm rot}^2 p_s/\left(G M_s(t)\right)$.  Also, let 
$J_0(t) \equiv J\left(\theta=0,t\right)$.  Then,

\[
P_z 
=
J_0(t) \left(k-1\right)
V_{\rm rot}
\]

\begin{equation}
\times \frac{\left(2T(t) - 5\right) \sqrt{2T(t) \left(1-T(t)\right)} + 3\sqrt{2} \arcsin{\sqrt{T(t)}} }
{16 T(t)}
,
\label{north1}
\end{equation}

\noindent{or}, in the small $T(t)$ approximation,

\begin{equation}
P_z \approx J_0(t) \left(k-1\right)
V_{\rm rot}
\left(
-\frac{1}{4\sqrt{2T(t)}}
+\frac{5\sqrt{T(t)}}{8\sqrt{2}}
\right)
,
\label{north2}
\end{equation}

\noindent{while} still $P_x = P_y = 0$.  Consequently,
in the adiabatic approximation, the eccentricity remains static.

\subsubsection{Longitudinal Dependence}

Here we explore the case of variation in mass loss with longitude.  
Let us assume the velocity dependence in equation (\ref{ueq}) holds, and
the dependence of $J$ on $\theta$ and $\phi$ are decoupled,
so that $J = J_0 j(\phi) j(\theta)$.  Here, $j(\theta) = 1 - T(t) \sin^2{\theta}$,
as in equation (\ref{Jeq}). Then,

\[
P_x =
\left[5 - 3 T(t) + \frac{3 \left(T(t)-1\right)^2 {\rm arctanh}{\sqrt{T(t)}}}{\sqrt{T(t)}} \right]
\]

\begin{equation}
\times \frac{1}{8} J_0(t) V_{\rm rot} \sqrt{ \frac{2}{T(t)} } 
\int_{0}^{2\pi} j(\phi) \cos{\phi} d\phi,
\end{equation}

\noindent{or}, in the small $T(t)$ approximation,

\begin{equation}
P_x \approx
J_0(t) V_{\rm rot}    
\left(\sqrt{\frac{2}{T(t)}} - \sqrt{2T(t)} \right)
\int_{0}^{2\pi} j(\phi) \cos{\phi} d\phi
,
\end{equation}

\noindent{and} similarly for $P_y$ except with $\cos{\phi}$ replaced
with $\sin{\phi}$. Also, $P_z = 0$.  Consequently, in both the general equations
of motion and the adiabatic approximation, a planet's inclination and longitude
of ascending node do not change owing to these longitudinal perturbations.

By analogy with the latitudinal case, let us consider when
the mass lost from the eastern hemisphere of the star is $k>1$ times
the amount of mass lost from the western hemisphere.  We assume
the eastern hemisphere is defined by the region bounded by
$\phi = 0$ and $\phi = \pi$.  Then
$P_x = 0$ and 

\[
P_y = \frac{k-1}{4} J_0(t) V_{\rm rot} \sqrt{ \frac{2}{T(t)} }  
\]

\begin{equation}
\times \left[5 - 3 T(t) + \frac{3 \left(T(t)-1\right)^2 {\rm arctanh}{\sqrt{T(t)}}}{\sqrt{T(t)}} \right] 
,
\label{east1}
\end{equation}

\noindent{or}, in the small $T(t)$ approximation,

\begin{equation}
P_y
\approx
J_0(t) V_{\rm rot} \left(k-1\right)
\left(\sqrt{\frac{8}{T(t)}} - \sqrt{8T(t)} \right)
.
\label{east2}
\end{equation}

\noindent{Although} this perturbation causes changes in the
planetary eccentricity, the argument of pericentre remains static.

\begin{table*}
 \centering
 \begin{minipage}{180mm}
 \centering
  \caption{Model parameters for Figs. \ref{k1}--\ref{k2}. The variable $t_f$ refers to the 
duration of the mass loss, $M_p$ the mass of the planet or other stellar companion, 
$M_s(0)$ the progenitor mass of the parent star, $M_s(t_f)$ the mass of the parent
star at time $t_f$, $p_s$ the parent star's radius, $V_{\rm rot}$ the rotational 
velocity of the star at the equator, $u(\theta)$ the mass outflow velocity
distribution, $u(\theta=0^{\circ})$ the polar mass outflow velocity,
$a(0)$ the initial semimajor axis, $e(0)$ the initial eccentricity, and 
$f(0)$ the initial true anomaly.}
  \begin{tabular}{@{}ccccccccccccc@{}}
  \hline
 Model \# & $\frac{t_f}{\rm Myr}$ & $\frac{M_p}{M_{\odot}}$ & $\frac{M_s(0)}{M_{\odot}}$ & $\frac{M_s(t_f)}{M_s(0)}$ & $\frac{p_s}{\rm au}$ & $\frac{V_{\rm rot}}{\rm km/s}$ & $\frac{V_{\rm rot}}{v_{\rm crit,max}}$ & $u\left(\theta\right)$ & $\frac{u\left(\theta=0^{\circ}\right)}{\rm km/s}$ & $\frac{a(0)}{\rm au} $ & $e(0)$ & $f(0)$\\
 \hline
 \hline
 \multicolumn{13}{|c|}{Adiabatic with Total Initial Mass = $1M_{\odot}$} \\ \hline
 1.1& 1.0 & 0 & 1.000 & 0.80 & 1.0 & -- & -- & flat & 50.0 & 35.0 & 0.25 & -- \\
 1.2& 1.0 & 0 & 1.000 & 0.70 & 1.0 & -- & -- & flat & 50.0 & 35.0 & 0.25 & -- \\
 1.3& 1.0 & 0 & 1.000 & 0.60 & 1.0 & -- & -- & flat & 50.0 & 35.0 & 0.25 & -- \\
 1.4& 1.0 & 0 & 1.000 & 0.50 & 1.0 & -- & -- & flat & 50.0 & 35.0 & 0.25 & -- \\
 1.5& 10.0 & 0 & 1.000 & 0.80 & 1.0 & -- & -- & flat & 50.0 & 35.0 & 0.25 & -- \\
 1.6& 10.0 & 0 & 1.000 & 0.70 & 1.0 & -- & -- & flat & 50.0 & 35.0 & 0.25 & -- \\
 1.7& 10.0 & 0 & 1.000 & 0.60 & 1.0 & -- & -- & flat & 50.0 & 35.0 & 0.25 & -- \\
 1.8& 10.0 & 0 & 1.000 & 0.50 & 1.0 & -- & -- & flat & 50.0 & 35.0 & 0.25 & -- \\
 1.9& 100.0 & 0 & 1.000 & 0.50 & 1.0 & -- & -- & flat & 50.0 & 35.0 & 0.25 & -- \\
 \hline
 \multicolumn{13}{|c|}{Adiabatic with Total Initial Mass = $2M_{\odot}$ and Total Final Mass = $0.8M_{\odot}$} \\ \hline
 2.1& 1.500 & 0.001 & 1.999 & 0.40 & 2.0 & -- & -- & flat & 0.1 & 4.00 & 0.1 & -- \\
 2.2& 1.500 & 0.001 & 1.999 & 0.40 & 2.0 & -- & -- & flat & 1.0 & 4.00 & 0.1 & -- \\
 2.3& 1.500 & 0.001 & 1.999 & 0.40 & 2.0 & -- & -- & flat & 10.0 & 4.00 & 0.1 & -- \\
 2.4& 1.500 & 0.001 & 1.999 & 0.40 & 2.0 & -- & -- & flat & 100.0 & 4.00 & 0.1 & -- \\
 2.5& 1.500 & 0.001 & 1.999 & 0.40 & 2.0 & -- & -- & flat & 0.1 & 15.00 & 0.8 & -- \\
 2.6& 1.500 & 0.001 & 1.999 & 0.40 & 2.0 & -- & -- & flat & 1.0 & 15.00 & 0.8 & -- \\
 2.7& 1.500 & 0.001 & 1.999 & 0.40 & 2.0 & -- & -- & flat & 10.0 & 15.00 & 0.8 & -- \\
 2.8& 1.500 & 0.001 & 1.999 & 0.40 & 2.0 & -- & -- & flat & 100.0 & 15.00 & 0.8 & -- \\
 2.9& 1.500 & 0.200 & 1.800 & 0.44 & 2.0 & -- & -- & flat & 10.0 & 15.00 & 0.8 & -- \\
 \hline
 \multicolumn{13}{|c|}{Non-adiabatic with Total Initial Mass = $2M_{\odot}$ and Total Final Mass = $0.8M_{\odot}$} \\ \hline
 2.11& 1.500 & 0.001 & 1.999 & 0.40 & 2.0 & 2.98 & 0.10 & equation (\ref{ueq}) & 42.11 & 1000 & 0.5 & $0^{\circ}$ \\
 2.12& 1.500 & 0.001 & 1.999 & 0.40 & 2.0 & 2.98 & 0.10 & equation (\ref{ueq}) & 42.11 & 1000 & 0.5 & $45^{\circ}$ \\
 2.13& 1.500 & 0.001 & 1.999 & 0.40 & 2.0 & 2.89 & 0.10 & equation (\ref{ueq}) & 42.11 & 1000 & 0.5 & $90^{\circ}$ \\
 2.14& 1.500 & 0.001 & 1.999 & 0.40 & 2.0 & 14.89 & 0.50 & equation (\ref{ueq}) & 42.11 & 1000 & 0.5 & $135^{\circ}$ \\
 2.15& 1.500 & 0.001 & 1.999 & 0.40 & 2.0 & 14.89 & 0.50 & equation (\ref{ueq}) & 42.11 & 1000 & 0.5 & $180^{\circ}$ \\
 2.16& 1.500 & 0.001 & 1.999 & 0.40 & 2.0 & 14.89 & 0.50 & equation (\ref{ueq}) & 42.11 & 1000 & 0.5 & $225^{\circ}$ \\
 2.17& 1.500 & 0.001 & 1.999 & 0.40 & 2.0 & 28.29 & 0.95 & equation (\ref{ueq}) & 42.11 & 1000 & 0.5 & $270^{\circ}$ \\
 2.18& 1.500 & 0.001 & 1.999 & 0.40 & 2.0 & 28.29 & 0.95 & equation (\ref{ueq}) & 42.11 & 1000 & 0.5 & $315^{\circ}$ \\
 2.19& 1.500 & 0.200 & 1.800 & 0.44 & 2.0 & 26.84 & 0.95 & equation (\ref{ueq}) & 39.96 & 1000 & 0.5 & $20^{\circ}$ \\
 \hline
\end{tabular}
\end{minipage}
\end{table*}

\subsubsection{Specific Examples}

We have shown (equations \ref{x0}--\ref{z0}) that the isotropic mass 
loss assumption is an excellent, physically-motivated approximation
to use for planetary motion in post-MS systems.  Now let us consider
the possibility that the stars do not behave according to 
equations (\ref{Jeq})--(\ref{ueq}).  Then, generally, $\vec{P} \ne 0$
and we may obtain relations such as equations (\ref{north1})--(\ref{north2}) or
 equations (\ref{east1})--(\ref{east2}).  The question we wish to answer is,
``How extremely asymmetric must giant star mass-loss be in order for the anisotropy 
to make a meaningful change in planetary orbit evolution?''
In order to conduct numerical simulations, 
we must first obtain physically-motivated values 
for $M_s$, $p_s$, $J_0$ and $V_{\rm rot}$.

We choose the mass of the star to be below that at which it would
end its life in a supernova so $M_s \lesssim 8M_{\odot}$ for
Solar metallicity and $M_s \lesssim 6M_{\odot}$ at ${\rm Z} = 10^{-4}$
\citep{huretal2000}.  Known exoplanet host stars nearly always 
have masses which are less than $3M_{\odot}$, although this mass barrier has
recently been broken \citep{satetal2012} and a few brown
dwarfs might orbit significantly more massive 
stars \citep[e.g.][]{hatetal2005}.  

The equatorial radius $p_s$ of the star is important 
because it expands dramatically during the post-MS. 
Often $p_s$ reaches a distance in au that is approximately
equal to the number of solar masses in the star's progenitor
mass (see e.g. Fig. 2 of \citealt*{veretal2013}).
Consequently, a planet may be engulfed by the expanding
envelope.  Even if a planet escapes this fate, the planet
may be affected by tidal forces from the envelope.  
The expansion also affects
the spin of the star, because conservation of angular momentum
dictates that stellar expansion in inversely correlated
with a star's rotational velocity.  We discuss this point
further in Section 3.1.4.

The other parameters, $J_0$ and $V_{\rm rot}$, are poorly
constrained observationally.  Additionally, much of the focus on 
stellar rotation theory centres around supernova progenitor 
stars with $M_s \gtrsim 8M_{\odot}$ \citep[e.g.][]{hegetal2000,eksetal2008,lovekin2011}, 
which are not our focus here.  Further, the time 
evolution of the rotational velocity of a post-MS star 
may be a strong function of the character of the mass 
loss (see, e.g., fig. 4 of \citealt*{maeder2002}).  To help
constrain our search for realistic values of $J_0$ and $V_{\rm rot}$, 
we consider two phases of post-MS evolution, the RGB and AGB phases.  

Consider first $J_0$.  For 
white dwarf progenitors with masses greater than about 
$2M_{\odot}$, the greatest combined mass loss occurs on the AGB 
(see e.g. fig. 7 of \citealt*{veretal2013}), and in 
particular at the tip of the AGB \citep[e.g.][]{vaswoo1993}.
Except for the AGB most mass loss typically occurs on the RGB.
The maximum mass-loss rate, however, nearly always occurs 
on the AGB, even accounting for the six different formulations of 
mass loss in RGB stars considered by \cite{catelan2009} 
[Reimers, Modified Reimers, Mullan, Goldberg, Judge-Stencel, 
and VandenBerg].
Empirical fits to observations \citep[e.g.][]{huretal2000}
demonstrate that, in some cases, stars with $M_s(t=0) \approx 8M_{\odot}$
can maintain mass-loss rates of the order of $10^{-4} M_{\odot}$yr$^{-1}$
for the order of $10^4$ yr.  These numbers may change 
to $10^{-6} M_{\odot}$yr$^{-1}$ and $10^4$ yr for some $1M_{\odot}$ stars,
and $10^{-8} M_{\odot}$yr$^{-1}$ and $10^7$ yr for other $1M_{\odot}$ stars 
\citep[see e.g.][]{verwya2012}.
During the period of greatest mass loss, at the tip of the AGB, 
$p_s$ is near its maximum value.

Now consider $V_{\rm rot}$.
The rotational velocities of stars are bounded from below by 
zero and from above by that equatorial velocity at
which a star would break up.  This critical
velocity is generally a function of the 
stellar luminosity.  \cite{maemey2000} investigate expressions 
for the critical velocity in detail; we are interested in the 
maximum of these values, so that we can have a representative 
range of potential rotation velocities to apply to our formulae.  
They show $v_{\rm crit, max} = \sqrt{2 G M_s/(3 p_{\rm polar})}$,
where $p_{\rm polar} = 2p_s/3$.  Hence, at the maximum critical
velocity $T = 1$, and the 
equations (\ref{north1}) and (\ref{east1}) become

\begin{equation}
P_{z}(T=1) = \frac{3\sqrt{2} \pi}{32} \left(k - 1\right) J_0 V_{\rm rot}
\approx 0.4 \left(k - 1\right) J_0 V_{\rm rot}
,
\end{equation}

\noindent{and}

\begin{equation}
P_{y}(T=1) = \frac{\sqrt{2}}{2} \left(k - 1\right) J_0 V_{\rm rot}
\approx 0.7 \left(k - 1\right) J_0 V_{\rm rot}
.
\end{equation}

\noindent{For} stars which do not rotate at all, $T=0$, and

\begin{eqnarray}
P_{z}(T=0) &=& -\frac{\left(k-1\right)}{4\sqrt{2}} J_0 \sqrt{\frac{GM_s}{p_s}}
\nonumber
\\
&\approx& -0.2 \left(k - 1\right) J_0 \sqrt{\frac{GM_s}{p_s}}
,
\end{eqnarray}

\begin{eqnarray}
P_{y}(T=0) &=& 2\sqrt{2} \left(k - 1\right) J_0 \sqrt{\frac{GM_s}{p_s}}
\nonumber
\\
&\approx& 2.8 \left(k - 1\right) J_0 \sqrt{\frac{GM_s}{p_s}}
.
\end{eqnarray}

All these considerations lead us to choose the parameters
presented in Table 1.  Although this work is not meant to represent 
an exhaustive phase space study, we have sampled a representative range 
of each parameter in the table.  In the first 18 models,
the secondary is close enough to the star such that the adiabatic approximation
may be used. In the last 9 models this approximation does not hold
and the general equations are used.  The secondary in the 
first 9 models roughly represents a Kuiper Belt Object such as Pluto.  
In the remaining models, the secondary mass is approximately equal to that of
either Jupiter or a $0.2M_{\odot}$ star.  In all models, 
we choose semimajor axes
and eccentricities such that the secondary's pericentre never
interacts with the stellar envelope nor approaches it 
closely enough that tides are likely
to become significant.  In their Fig. 7, \cite{musvil2012} demonstrate
that for low eccentricity orbits ($e \lesssim 0.2$) the maximum 
planet radius for which a planet becomes tidally
engulfed is located at or within the maximum stellar radius achieved 
for terrestrial-mass or Neptune-mass planets, but not 
Jovian-mass planets.  The Jovian-mass planets may be enveloped at distances of 
approximately $3.5$\,au or $2.7$\,au if they orbit stars with
progenitor masses that are approximately equal to $2 M_{\odot}$
and $1 M_{\odot}$, respectively.  However, tidal effects for 
moderately and highly eccentric cases ($e \gtrsim 0.2$) are
poorly constrained, because tidal theories often rely on low-eccentricity
expansions.

\begin{figure}
\centerline{
\psfig{figure=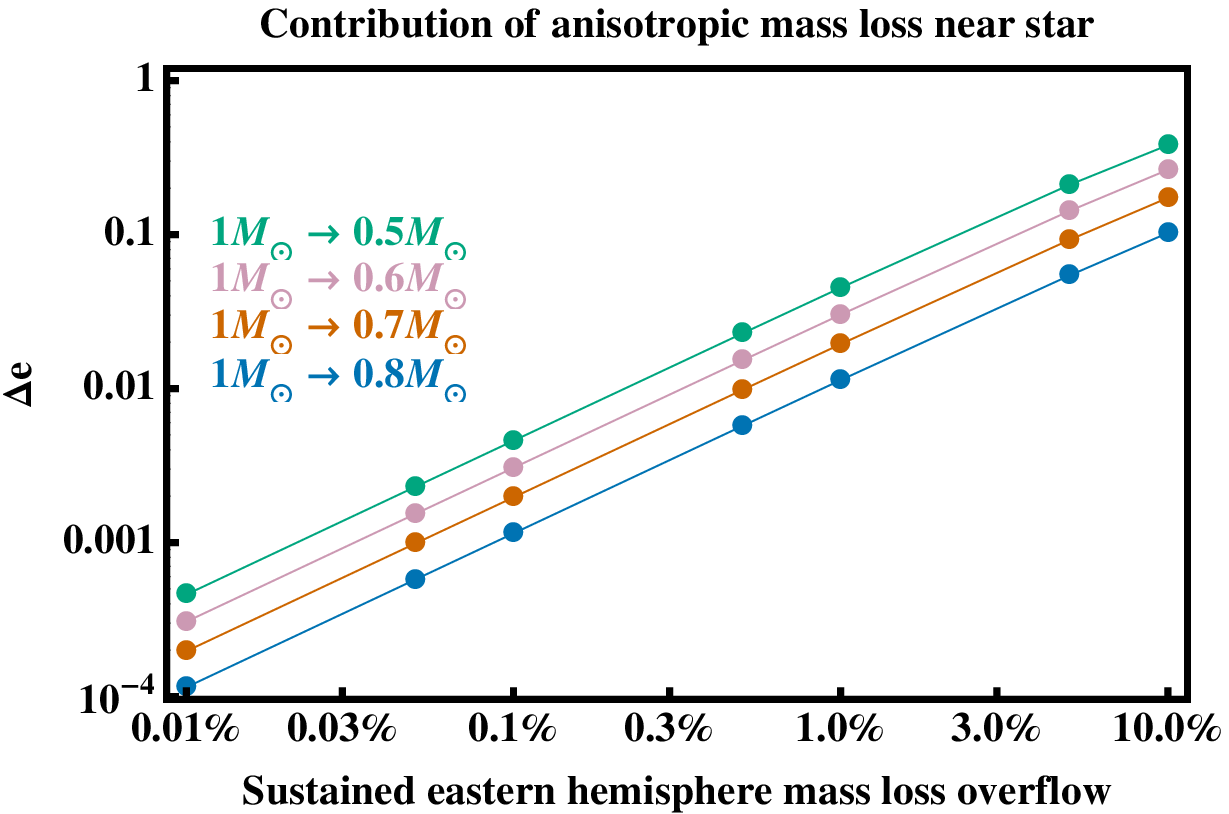,height=6.0cm,width=9.0cm} 
}
\centerline{
\psfig{figure=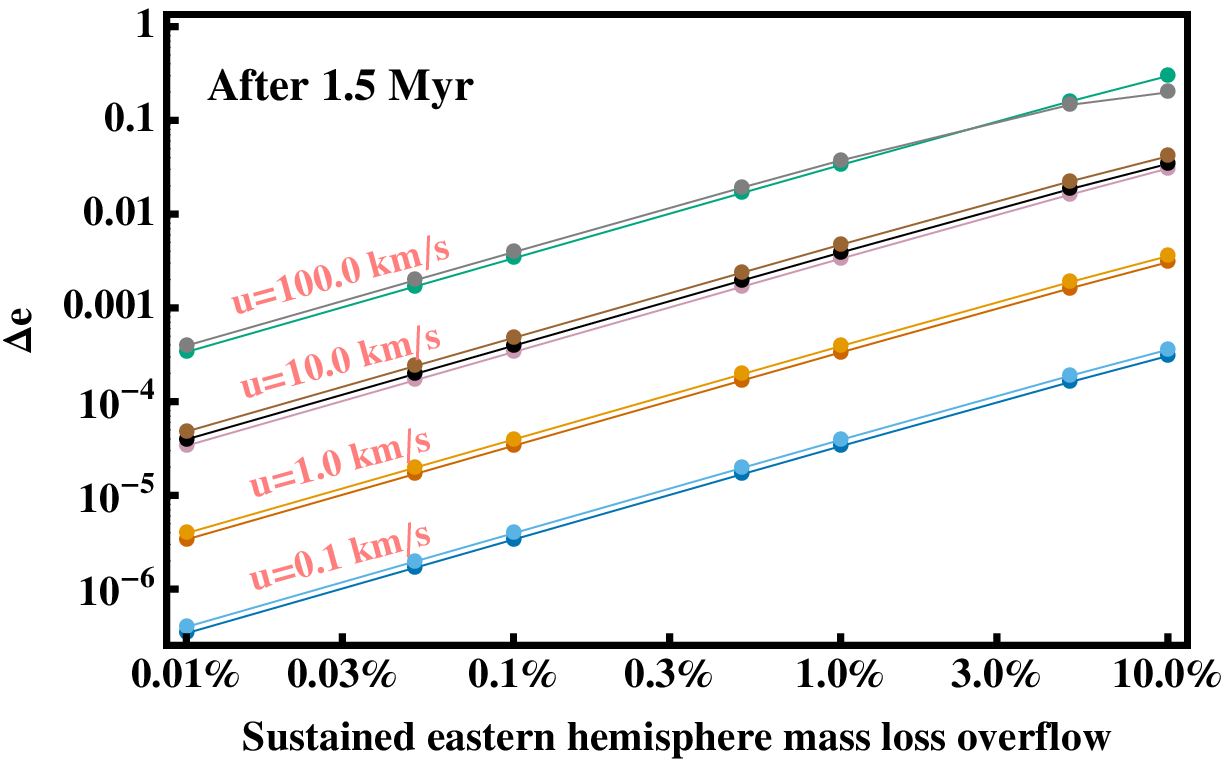,height=6.0cm,width=9.0cm} 
}
\caption{
Assessing the robustness of the isotropic mass-loss
approximation in post-MS studies.  Plotted is
the adiabatic change in a secondary's eccentricity versus
$k-1$ due to the excess mass fraction lost from the eastern
hemisphere.  We use the anisotropic mass-loss prescription
of equation (\ref{east1}).  In the upper panel, the blue
(bottom) curve represents models 1.1 and 1.5 from Table 1.  
Moving up, the orange curve represents models 1.2 and 1.6,
the purple curve models 1.3 and 1.7, and the green
curve models 1.4, 1.8 and 1.9.  In the lower panels,
the curves from bottom to top represent models 2.1,
2.5, 2.2, 2.6, 2.3, 2.7, 2.9, 2.4 and 2.8.  These plots
demonstrate that anisotropic mass loss must be sustained
at least at the $1$ per cent level over at least 1 Myr to produce
an observable difference from the isotropic mass-loss case.}
\label{k1}
\end{figure}

\begin{figure}
\centerline{
\psfig{figure=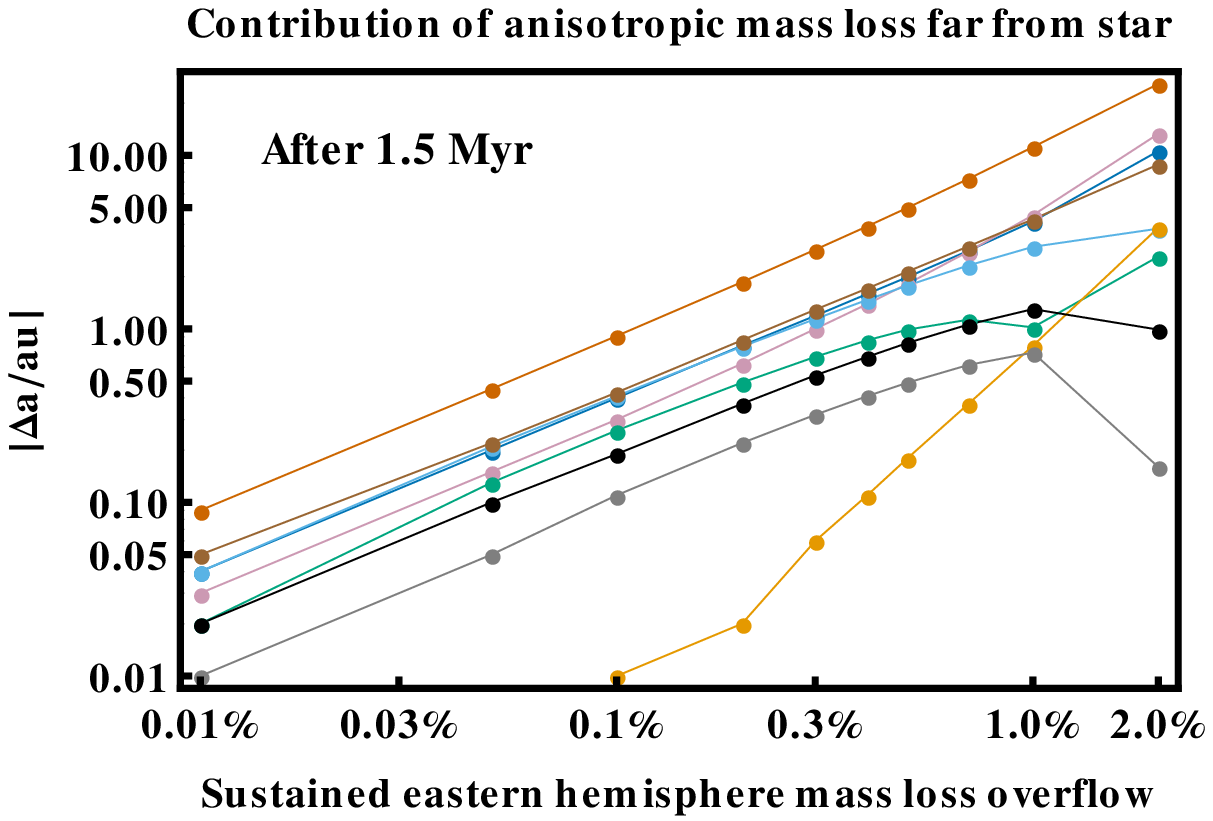,height=6.0cm,width=9.0cm} 
}
\centerline{
\psfig{figure=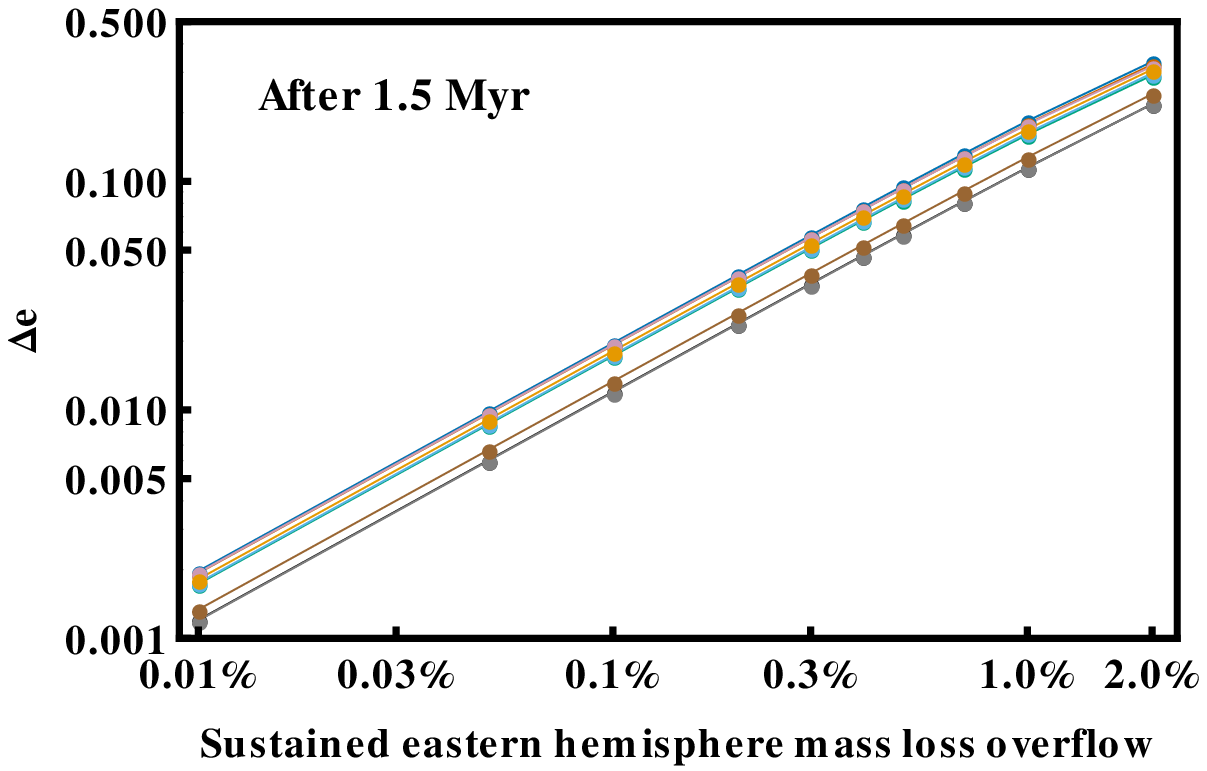,height=6.0cm,width=9.0cm} 
}
\caption{
Like Fig. \ref{k1} but for the non-adiabatic case,
where the secondary is at $a(0) = 10^3$ au.  Anisotropic
mass-loss has a greater effect in the outer reaches
of planetary systems.
}
\label{k2}
\end{figure}

The quantity $t_f$ present in the table represents the total time
during which a star experiences $k$ times as much mass loss from
the eastern hemisphere as the western hemisphere, as in equation (\ref{east1}).
We choose this scenario because (1) in the 
physically-motivated scenario, where both $k=1$
and $u$ satisfies equation (\ref{ueq}), the anisotropic terms make zero
contribution to the planet's motion.  This result demonstrates the excellence
of the isotropic mass-loss approximation.  Therefore, we must somehow
depart from this scenario.  (2) By instead considering a north-south dichotomy
in the star, as in equation (\ref{north1}), we would not be able to quantify
the change in $a$ or $e$ due to the anisotropic terms because 
$P_x = P_y = 0$ (see equations \ref{dadtgenP}--\ref{dedtgenP}).
(3) The adiabatic equations (equations \ref{dadtP}--\ref{dedtP}) 
admit a closed, analytical solution for $e(t)$ if we
assume that both the mass-loss rate and ejecta velocity are constant.  
Because the solution is similar to equations (\ref{at1})--(\ref{et2}), but with
a dependence on $k$ and Heaviside step functions, the equations and 
numerical simulations can be used to check each other.  Subsequently,
the analytical equations may be used to quickly model over $10^5$ 
orbits of a planetary companion, such as models 2.1--2.4. 
(4) Anisotropic behaviour is likely to be transient and averaged
out given enough time.  Therefore, by presenting an extreme
scenario such as sustained mass loss from one hemisphere, we
are presenting an upper bound on the effects of anisotropy.

Consider first the 18 adiabatic models (models 1.1--2.9).  In the 
adiabatic isotropic mass
loss case $e$ remains fixed and $a$ varies, whereas in the
anisotropic case $e$ varies and $a$ incurs no additional variation.  
Therefore, the eccentricity
achieved in the anisotropic case is due entirely to the anisotropic
terms.  Consequently, in Fig. \ref{k1}, we plot $\Delta e$ as
a function of $k-1$ for models 1.1--1.9 (upper panel) and models
2.1--2.9 (lower panel).  Dots represent the outcome of numerical 
simulations for specific values of $k$.  In the upper panel,
we find that the different values of $t_f$ chosen make no 
discernible difference to the outcome for the same total
mass loss.  In other words, models
1.1 and 1.5 yield the same result, as do models 1.2 and 1.6,
1.3 and 1.7, and 1.4 and 1.8 and 1.9.  In the lower panel
all 9 models produce distinct curves on the plot.  However,
these are sharply grouped according to ejecta velocity.
One order of magnitude difference in a uniform ejecta velocity
roughly corresponds to one order of magnitude difference
in the secondary's eventual eccentricity.  In contrast, 
when the secondary is a small star (brown curve, third from top), 
with a mass about two orders of magnitude greater than Jupiter's mass,
the eccentricity increases by just a few tens of percent.
Similarly, a 0.7 change in the initial eccentricity makes
little difference relative to changes in $u$.  The primary
difference is the kink seen for the grey curve ($e_0 = 0.8$) 
at $k=1.100$; as the eccentricity
approaches unity, $de/dt$ slows.

Now consider the 9 non-adiabatic models (2.11--2.19).
In this case, anisotropy affects both $a$ and $e$,
the upper and lower panels of Fig. \ref{k2}. The behaviour is more complex
in the general case because of the dependence on $f$.
This dependence in the isotropic case is discussed in detail 
by \cite{veretal2011} and \cite{verwya2012}.  Here we are 
just interested in the change due to anisotropy.
The upper plot shows non-monotonic curves,
as well as instances when $\Delta a$ may reach about $1$ per cent
of $a(0)$.  The magnitude of the eccentricity change for these
$a(0) = 1000$ au models for a given $k$ is greater than for the 
corresponding models in Fig.~\ref{k1}.  This comparison 
corroborates the effect of the extra factor of $\sqrt{a}$ 
in the anisotropic terms.  Nevertheless, the eccentricity 
still appears to increase uniformly, as in the adiabatic case, 
despite the introduction of latitudinal dependence on $u$.
These plots confirm the finding of \cite{paralc1998} that
anisotropic mass loss is important primarily in the outer
reaches of planetary systems.

Overall, these scenarios quantify the extent
to which sustained, strong anisotropic
mass loss is necessary in order to produce an observable change
in a secondary's orbital parameters.

\subsubsection{Complex behaviour of giant stars}

We have demonstrated that giant-star anisotropic mass loss
may be a strong function of $V_{\rm rot}$.
Our giant star applications in this work 
established bounds on the strength of the
asymmetry with simple assumptions
about rotation and mass flux.  In reality,
these attributes are complex and time dependent,
but largely unknown.  

In particular, arguments appealing to the conservation 
of angular momentum in AGB stars -- which would dramatically spin down 
the expansive envelope -- largely fail to fit with the 
dominant fraction of observed planetary nebulae that are 
aspherical \citep{kimetal2008}.  Additionally, 
despite observations that rapidly rotating AGB stars
are classified as very rare \citep{perroi2006},
one exception, that of V Hya \citep{baretal1995}, rotates at nearly critical 
velocity.  This rapid rotation may be explained by 
a binary origin for the star.  Another exception mentioned is ZNG 1, which 
has a measured rotational velocity of $170 \pm 20$ 
km s$^{-1}$ \citep{dixetal2004}.  \cite{dijspe2006} discuss the 
sources and sinks of angular momentum which can break the 
assumption of conservation.  Although differential 
rotation has been observed in giant stars \citep{mosetal2012},
observations cannot yet pinpoint the particular
form of the angular momentum distribution inside horizontal 
branch stars.  See \cite{silpin2000} for four 
potential angular momentum distributions.  Also, 
see fig. 7 of \cite{heglan1998} 
for a sample of rotation profiles due to different assumptions 
about the internal structure of supergiants.

A complicating factor for mass flux is how the mass loss changes owing to
thermal pulses on the AGB. One of the few observational examples of AGB
mass loss is due to the spectacular shell of dust and gas
around R Sculptoris \citep{maeetal2012}.  This shell was created
from ejecta from a single thermal pulse.  The authors estimate that the
pre-pulse, intra-pulse, and post-pulse mass-loss rates over about 200 yr were 
on the order of, respectively, $10^{-6}M_{\odot}$yr$^{-1}$, 
$10^{-5}M_{\odot}$yr$^{-1}$, and $10^{-7}M_{\odot}$yr$^{-1}$.  Hence, 
the observations indicate a difference of about 2 orders of magnitude 
in mass loss on the order of $100$ yr for just a single pulse.

The overall lesson from these studies is that, 
in order to model individual systems where
observations fail to provide constraints, 
one should consider a wide variety
of parameters.

\subsection{Birth Jets}

Currently observed exoplanets orbiting MS stars may have been
affected by perturbations due to stellar outflows during their formation.  
In this section, we quantify how 
a planet's orbital parameters may have changed.

\subsubsection{Context and Assumptions}

Unlike in the post-MS case, where mass loss alone affects the orbit of a planet
outside of the tidal reach of the parent star, in the pre-MS case multiple entities
may perturb a planet.  However, identifying the potential presence and influence of
the protoplanetary disc, accretion on to the star, and the release valves
for outflows {\it after} a planet has been formed from that disc is challenging.

The transition between the end of star formation \citep[e.g.][]{hartmann1998,mckost2007}
and the beginning of planet formation \citep[e.g.][]{armitage2010} is murky, and both
processes are not mutually exclusive.  Only
a handful of authors have attempted to bridge this gap \citep[e.g.][]{larson2002,griv2007}
although there exists an acknowledgment that the presence of planets can
affect the accretion rate on to the star \citep[e.g.][]{gooraf2001,naylod2012,nayakshin2013}.
Therefore, questioning whether a young system with bipolar outflows can
plausibly host a fully-formed planet is valid.  The formation of bipolar jets themselves
\citep{blapay1982,pudnor1983} is a hallmark of accretion on to protostars, and hence 
a disc often, but not necessarily always\footnote{\cite{schetal2013} describe 
potential disc-based and protoplanet-based triggers of accretion in their introduction.}, accompanies jets. 
The proportion of the disc which is dissipated due to accretion, photoevaporation and 
planet formation, as a function of time,
remains unknown both observationally and theoretically.

Nevertheless, this picture may be constrained.  In some cases, 
accretion rates on to the star as high as $5 \times 10^{-7} M_{\odot}$yr$^{-1}$ can persist
for $1$ Myr, and accretion rates as high as $10^{-8} M_{\odot}$yr$^{-1}$ can persist
for $10$ Myr, after the star is born \citep[fig. 7 of][]{armetal2003}.  
These lifetimes are commensurate with those of protoplanetary discs
\citep[fig. 2 of][]{wyatt2008}, which place hard constraints on the timescale for planet
formation.  Stars that react to accretion from discs by expelling matter do so 
through X-winds \citep[e.g.][]{shuetal1994,shuetal2000,shaetal2007} at inner
disc edges \citep{cai2009}, which typically reside at hundredths of au, or through disc-winds 
\citep{konpud2000,pudetal2007} at tenths of an au to several au \citep[e.g.][]{andetal2003}.
The mass accretion rate is typically one order of magnitude higher than
than the mass ejection rate \citep{cabetal1990,haretal1995,hartmann1998,pudetal2007}. 

The dual processes of mass accretion and mass ejection may both strongly affect
an orbiting planet.  However, the ejected mass is being ejected from the system,
whereas the accreted mass is being transposed within the system.  Therefore, 
due to accretion, a planet which is external to the accretion disc shifts 
its orbit only to the extent that the central star has effectively changed its 
moment of inertia.  The planet's osculating Keplerian parameters are not
able to undergo the major changes which can accompany mass loss from the system.
The further away the planet resides from the disc, the smaller the effect from
accretion.  Henceforth, we neglect mass accretion in our computations.

The disc itself provides another perturbation on the planet (and vice versa).
\cite{namouni2005} and \cite{namouni2013}, who also consider the effect on 
planets of bipolar jets, neglect the disc mass.  We similarly neglect
the disc, but first provide some more detailed justification for doing so.
Protoplanetary discs, which are gas-poor remnants of protostellar discs, have
observed disc masses ranging from, for example, 
$10^{-3}$--$10^{-1}M_{\odot}$ \citep[fig. 1 of][]{wilcie2011}.

Given these typical disc masses, we can compute the ratio of the 
gravitational potential, $\psi$, of the disc to the total
potential (from the disk and star) at arbitrary planet locations.
We demonstrate that this ratio is 
small.  We utilize the general potential formulae of \cite{hure2012},
and consider two cases.  The first is when the planet is coplanar to the disk and
the second is when the planet-star line is perpendicular to the disc.  
The latter occurs in the extreme
case of a polar orbit.  The disc is assumed to be a thin circular annulus
extending from $0.05$ au to an arbitrary distance $R_{d}'$.  Then

\[
\psi_{\rm disk,coplanar} = \psi(R_{d}^{(c)} = R_d') - \psi(R_{d}^{(c)} = 0.05 \ {\rm au})
,
\nonumber
\]

\[
\psi_{\rm disk,vertical} = \psi(R_{d}^{(v)} = R_d') - \psi(R_{d}^{(v)} = 0.05 \ {\rm au})
\nonumber
,
\]

\noindent{where}

\[
\psi(R_{d}^{(c)}) = -2G\Sigma 
\nonumber
\]

\[
\times \Bigg[ 
\left(R_{d}^{(c)} + R_p\right) 
\int_{0}^{\frac{\pi}{2}}  
\bigg(  
1
-
\frac{4R_{d}^{(c)}R_p}{\left(R_{d}^{(c)} + R_p\right)^2}
\sin^2{\varsigma}
\bigg)^{\frac{1}{2}}
 d\varsigma
\]

\begin{equation}
+
\left(R_{d}^{(c)} - R_p\right)
\int_{0}^{\frac{\pi}{2}}  
\bigg(  
1
-
\frac{4R_{d}^{(c)}R_p}{\left(R_{d}^{(c)} + R_p\right)^2}
\sin^2{\varsigma}
\bigg)^{-\frac{1}{2}}
 d\varsigma
\Bigg]
,
\end{equation}

\begin{equation}
\psi(R_{d}^{(v)}) = -2 \pi G \Sigma \left[\sqrt{R_{d}^{{(v)}^2} + R_{p}^2} - R_{d}^{(v)}  \right]
\end{equation}

\begin{figure}
\centerline{
\psfig{figure=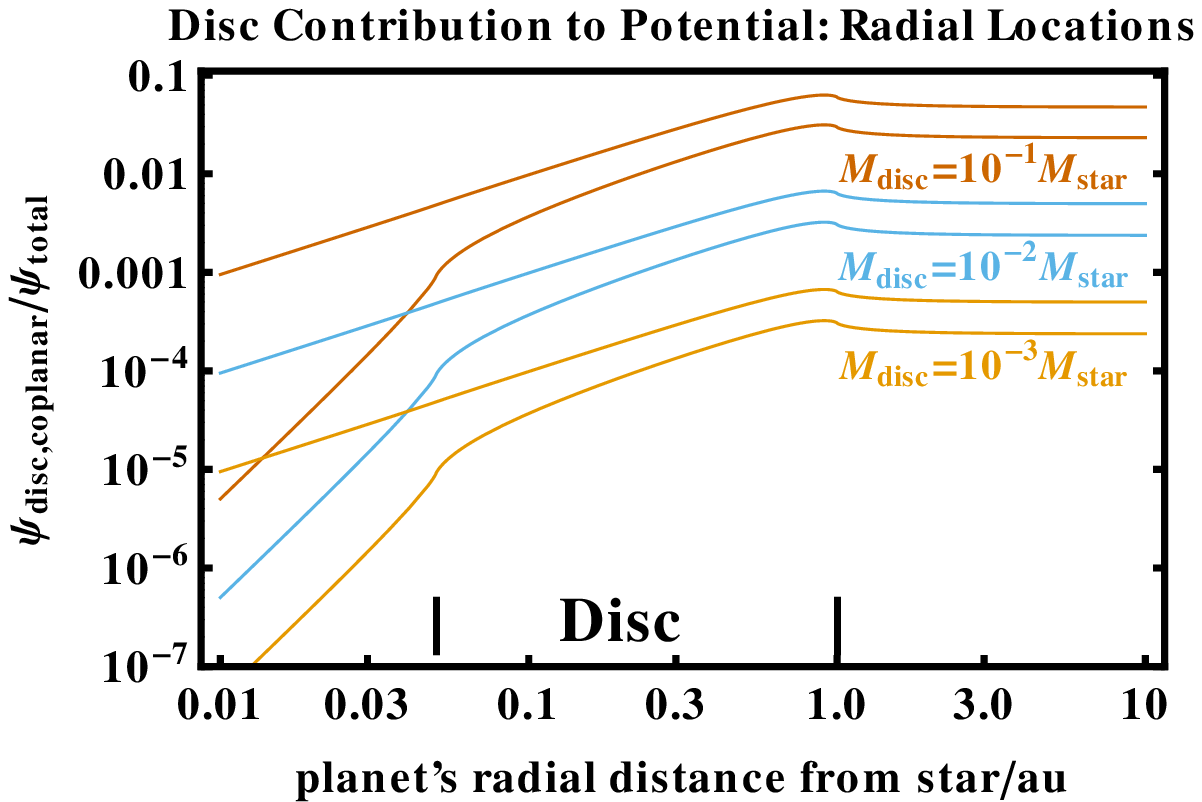,height=6.0cm,width=9.0cm} 
}
\centerline{}
\centerline{
\psfig{figure=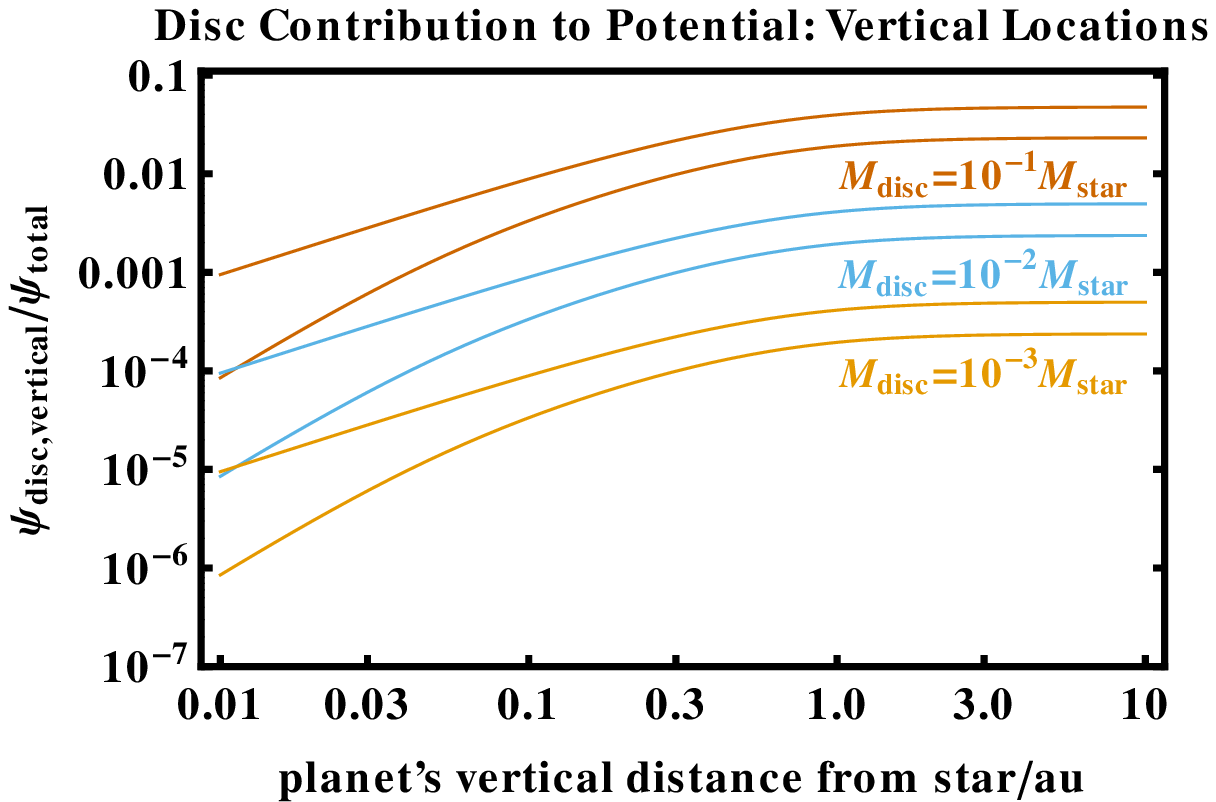,height=6.0cm,width=9.0cm} 
}
\caption{
A demonstration of why accretion discs may be neglected
for nascent system mass loss studies.  The
top and bottom plots show the fraction of the total
gravitational potential on a planet due to a thin disc
annulus from 0.05 - 1.0 au for radial and polar 
planet locations, respectively.
For each pair of curves, the top and bottom 
assume $r^{-1}$ and $r^{-3/2}$ mass distributions
in the disc.}
\label{discpert}
\end{figure}

\noindent{and} $\Sigma$ is the disc surface density,
which we assume to be a power law.  In Fig. \ref{discpert},
we plot the results for $R_{d}' = 1$ au and two power-law
exponents.  For each pair of similarly-coloured curves,
the top represents the Keplerian
power law exponent ($-3/2$) and the bottom represents
a linear exponent ($-1$).

For all curves, the disc contribution to the potential
never reaches a tenth, and achieves a hundredth only
when $M_{\rm disc} > 0.01 M_s$.  Both cases presented are broadly
similar, and do include feedback on the disc.  As 
expected, the disc contribution becomes asymptotic for 
large $R_p$ and approaches zero as $R_p \rightarrow 0$.  
Because of this asymptotic behaviour, when modelling an
external planet affected by jet-induced mass ejection, one 
may simply add the stellar mass to the disc mass and treat both 
as a single entity.

In summary, we believe the assumptions of \cite{namouni2013}
of neglecting accretion and influences from the disc
are valid for the precision sought.  We make the
same assumptions in the subsequent analysis.

\subsubsection{General Jets}

In Appendix A, we consider the movement of a planet caused by 
a jet of material emanating from the star in an 
arbitrary direction.  The equations there also incorporate
the mass loss expected from the rest of the star along
with the jet.

\subsubsection{Asymmetric Bipolar Jets}

\begin{table*}
 \centering
 \begin{minipage}{180mm}
 \centering
  \caption{Model parameters for Figs. \ref{j1}-\ref{j2}.  Here, $\dot{M}_{\rm up}$,
$\dot{M}_{\rm down}$, $u_{\rm up}$ and $u_{\rm down}$ refer to the mass loss rate 
and outflow velocity at the north pole ($\theta = 0$, ``up'') and 
south pole ($\theta = \pi$, ``down'').}
  \begin{tabular}{@{}ccccccccc@{}}
  \hline
 Model \# & $\frac{M_p}{M_{\odot}}$ & $\frac{M_s(0)}{M_{\odot}}$ & $\frac{\dot{M}_{\rm up}}{M_{\odot}/{\rm yr}}$ & $\frac{\dot{M}_{\rm down}}{M_{\odot}/{\rm yr}}$ & $\frac{u_{\rm up}}{\rm km/s}$ & $\frac{u_{\rm down}}{\rm km/s}$ & $\frac{a(0)}{\rm au} $ & $e$ \\
 \hline
 \hline
 \multicolumn{9}{|c|}{Adiabatic with Total Initial Mass = $1M_{\odot}$} \\ \hline
 J1.1 & 0.001 & 0.999 & $1\times10^{-10}$ & $2\times10^{-10}$ & 10 & 20 & 5.0 & 0.5 \\
 J1.2 & 0.001 & 0.999 & $1\times10^{-10}$ & $2\times10^{-10}$ & 100 & 200 & 5.0 & 0.5 \\
 J1.3 & 0.001 & 0.999 & $1\times10^{-10}$ & $4\times10^{-10}$ & 100 & 200 & 5.0 & 0.5 \\
 J1.4 & 0.001 & 0.999 & $1\times10^{-8}$ & $2\times10^{-8}$ & 10 & 20 & 5.0 & 0.5 \\
 J1.5 & 0.001 & 0.999 & $1\times10^{-8}$ & $2\times10^{-8}$ & 100 & 200 & 5.0 & 0.5 \\
 J1.6 & 0.001 & 0.999 & $1\times10^{-8}$ & $4\times10^{-8}$ & 100 & 200 & 5.0 & 0.5 \\
 J1.7 & 0.001 & 0.999 & $1\times10^{-6}$ & $2\times10^{-6}$ & 10 & 20 & 5.0 & 0.5 \\
 J1.8 & 0.001 & 0.999 & $1\times10^{-6}$ & $2\times10^{-6}$ & 100 & 200 & 5.0 & 0.5 \\
 J1.9 & 0.001 & 0.999 & $1\times10^{-6}$ & $4\times10^{-6}$ & 100 & 200 & 5.0 & 0.5 \\
 \hline
 \multicolumn{9}{|c|}{Adiabatic with Total Initial Mass = $2M_{\odot}$} \\ \hline
 J2.1 & 0.001 & 1.999 & $4\times10^{-7}$ & $5\times10^{-7}$ & 50 & 75 & 1.0 & 0.01 \\
 J2.2 & 0.001 & 1.999 & $4\times10^{-7}$ & $5\times10^{-7}$ & 50 & 75 & 1.0 & 0.2 \\
 J2.3 & 0.001 & 1.999 & $4\times10^{-7}$ & $5\times10^{-7}$ & 50 & 75 & 1.0 & 0.7 \\
 J2.4 & 0.001 & 1.999 & $4\times10^{-7}$ & $5\times10^{-7}$ & 50 & 75 & 3.0 & 0.01 \\
 J2.5 & 0.001 & 1.999 & $4\times10^{-7}$ & $5\times10^{-7}$ & 50 & 75 & 3.0 & 0.2 \\
 J2.6 & 0.001 & 1.999 & $4\times10^{-7}$ & $5\times10^{-7}$ & 50 & 75 & 3.0 & 0.7 \\
 J2.7 & 0.001 & 1.999 & $4\times10^{-7}$ & $5\times10^{-7}$ & 50 & 75 & 10.0 & 0.01 \\
 J2.8 & 0.001 & 1.999 & $4\times10^{-7}$ & $5\times10^{-7}$ & 50 & 75 & 10.0 & 0.2 \\
 J2.9 & 0.001 & 1.999 & $4\times10^{-7}$ & $5\times10^{-7}$ & 50 & 75 & 10.0 & 0.7 \\
 \hline
\end{tabular}
\end{minipage}
\end{table*}

\begin{figure}
\centerline{
\psfig{figure=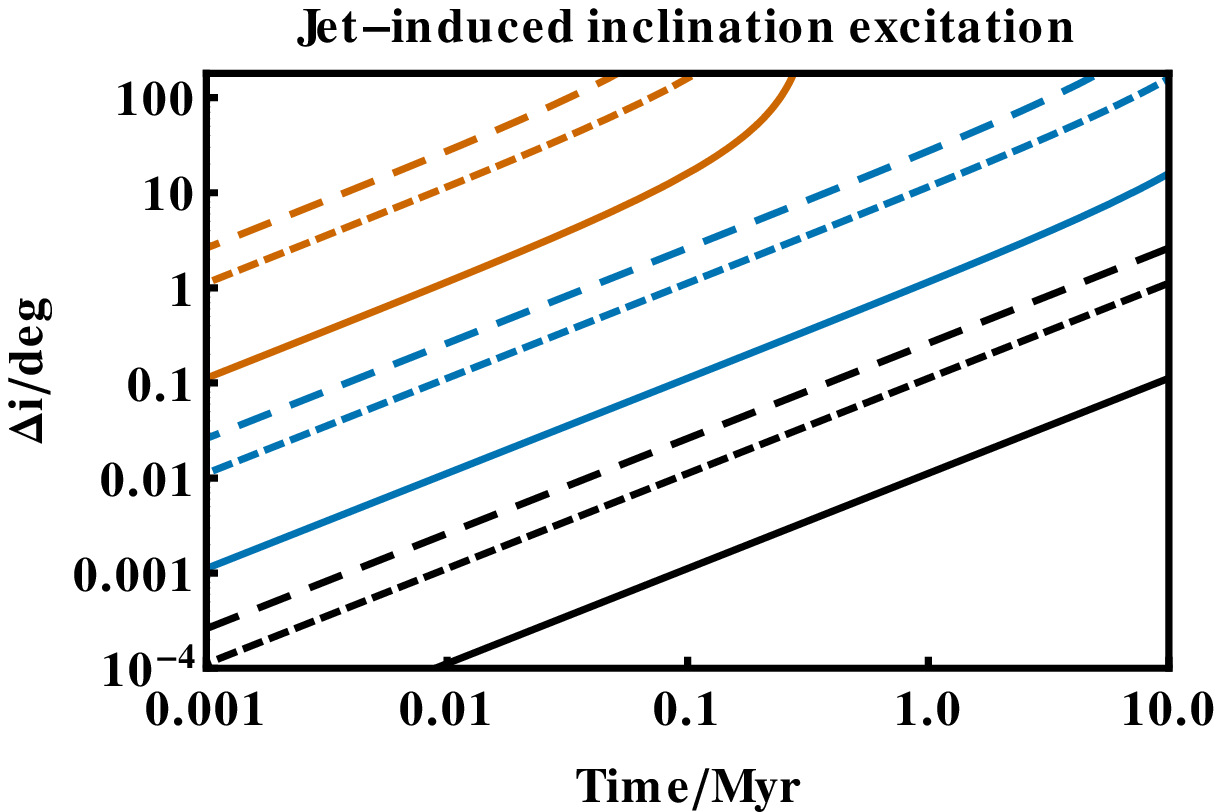,height=6.0cm,width=9.0cm} 
}
\centerline{
\psfig{figure=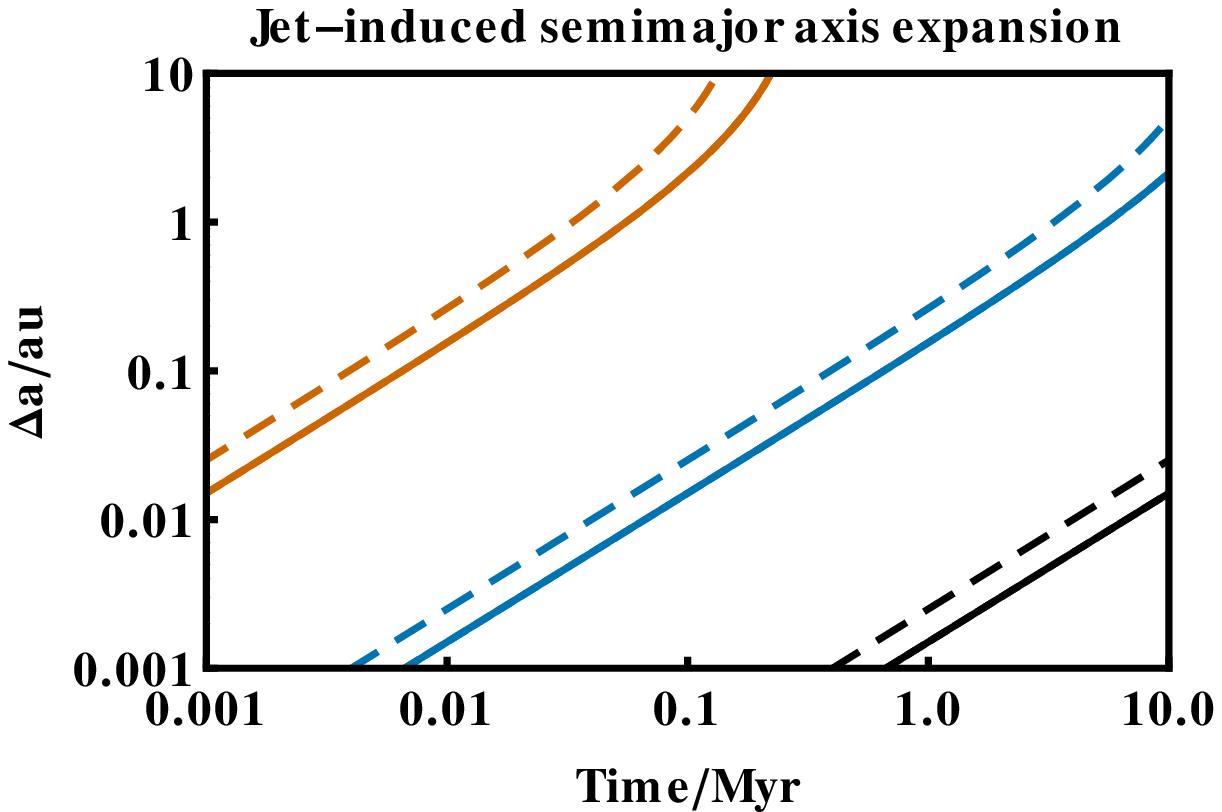,height=6.0cm,width=9.0cm} 
}
\caption{
How asymmetric bipolar jets tilt and expand an orbit.
The plots illustrate the change in the planet's 
inclination ({\it upper panel}) and semimajor 
axis ({\it lower panel}) in models J1.1-J1.9
in Table 2.  The model number increases monotonically
from the bottom curve to the top in the upper panel.
In the lower panel the solid and short-dashed curves
are equivalent. Curves of the same colour correspond
to roughly the same order of magnitude for mass-loss
rates.  The plots demonstrate that a realistic
sample of models can reproduce observed
exoplanet inclinations.}
\label{j1}
\end{figure}

\begin{figure}
\centerline{
\psfig{figure=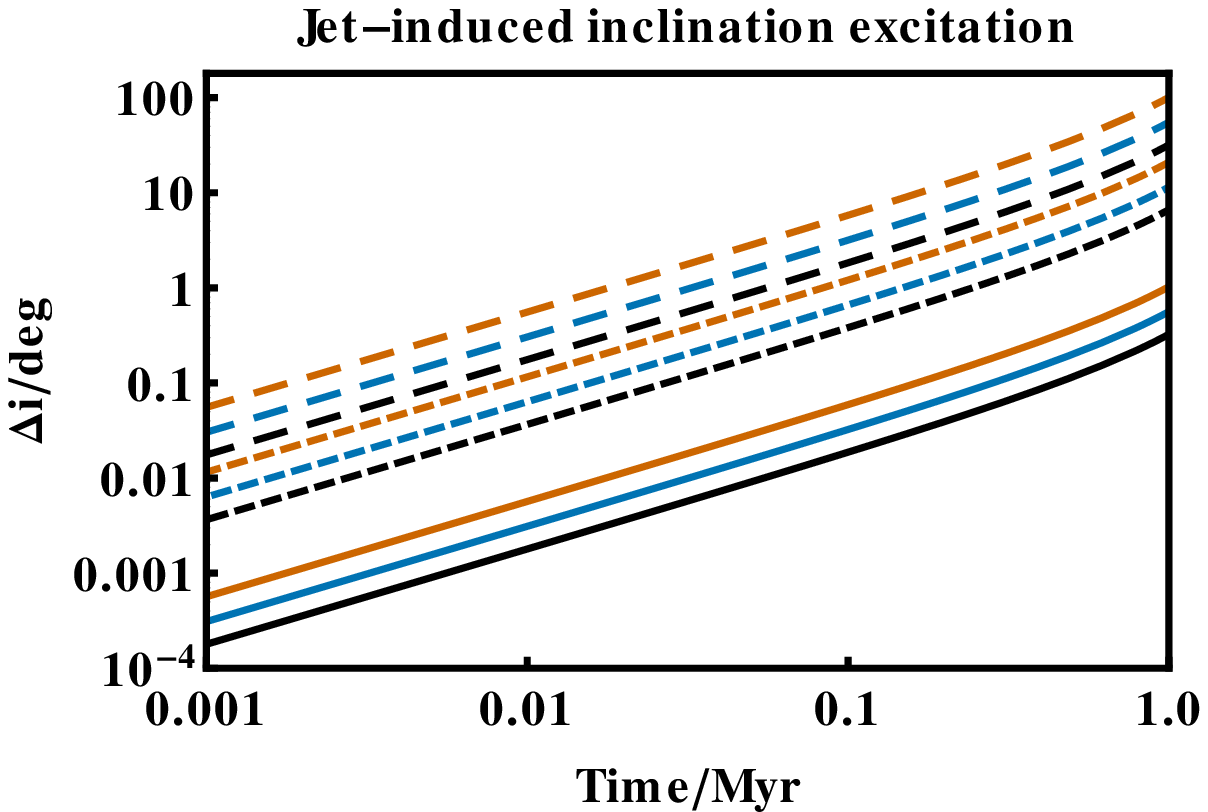,height=6.0cm,width=9.0cm} 
}
\centerline{
\psfig{figure=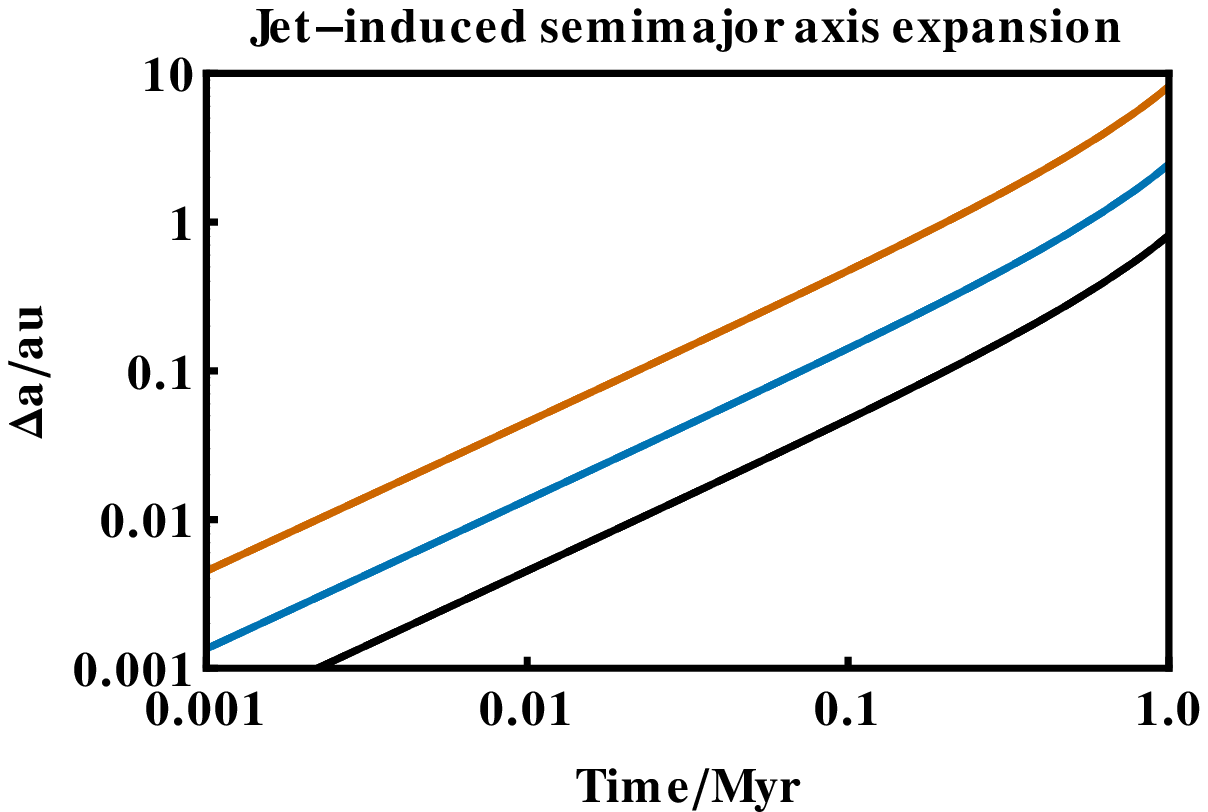,height=6.0cm,width=9.0cm} 
}
\caption{
The same as Fig. \ref{j1} but for models
J2.1-J2.9 in Table 2. There include variations
in the initial planetary orbit.  In the upper panel,
starting at the bottom curve and moving up,
the curves correspond to models J2.1, J2.4,
J2.7, J2.2, J2.5, J2.8, J2.3, J2.6 and J2.9.
Curves of the same colour correspond to the
same value of $a(0)$.  Hence, because 
the equivalently-coloured curves are so widely separated, 
the upper panel demonstrates the strong dependence 
of $i(t)$ on $e$.}
\label{j2}
\end{figure}

One physically-motivated case of particular interest is bipolar jets.  In the limit of 
infinitesimally thin jets which contain all the stellar mass loss, there is no longitudinal 
dependence on mass-loss, and hence $P_x = P_y = 0$.  Therefore 
$F_x = \mathsf{Q}_{13}P_z$, $F_y = \mathsf{Q}_{23}P_z$ and 
$F_z = \mathsf{Q}_{33}P_z$.  When both jets are completely symmetric, they eject
mass at the same rate and velocity, so there is no net anisotropic force on the planet 
{\it regardless of the planet's location}.

Consider ideal asymmetric bipolar jets which do produce a net anisotropic force on a planetary orbit.  
We can deduce several properties of the resulting motion just by considering equations (\ref{dadtgenP})--(\ref{domdtgenP})
and equations (\ref{dadtP})--(\ref{domdtP}), with $P_x = P_y = 0$.  
The reduced form of equations (\ref{dadtP})--(\ref{domdtP}) has the following properties.

\begin{itemize}
\item{1) A planet's eccentricity remains fixed, just as in the isotropic
adiabatic mass-loss case.} 
\item{2) An eccentric planet has an argument of pericentre that 
precesses continuously unless $\omega_0 = 0$.}
\item{3) The magnitudes of $di/dt$, $d\Omega/dt$ and $d\omega/dt$ all correlate positively with eccentricity. }
\end{itemize}

We analyze the case $\omega_0 = 0$ in the adiabatic regime 
because this case admits complete analytical 
solutions for the orbit.  The resulting motion demonstrates 
how the semimajor axis and inclination evolve with a static
eccentricity.  Let us denote the mass-loss rate and velocity 
at $\theta = 0$ and $\theta = \pi$
as $\lbrace \dot{M}_{\rm up}>0, u_{\rm up}>0\rbrace$ and 
$\lbrace \dot{M}_{\rm down}>0, u_{\rm down}>0\rbrace$, and treat
these quantities as constant in time.  Hence, we solve

\begin{eqnarray}
\frac{da}{dt} &=& -\frac{a(t)}{M_s(t) + M_p}
\left(\dot{M}_{\rm up} + \dot{M}_{\rm down} \right)
\label{jeteq1}
,
\\
\frac{di}{dt}
&=&
\frac{3e}{2 M_s(t)\sqrt{1 - e^2}} \sqrt{\frac{a(t)}{G\left(M_s(t) + M_p\right)}}
\nonumber
\\
&\times& 
\left( \dot{M}_{\rm down} u_{\rm down} - \dot{M}_{\rm up} u_{\rm up} \right)
.
\label{jeteq2}
\end{eqnarray}

\noindent{Recall} that the eccentricity is constant.  For a finite secondary mass, the
solutions to equations (\ref{jeteq1})--(\ref{jeteq2}) are

\begin{eqnarray}
M_s(t) &=& M_s(0) - t \left(\dot{M}_{\rm up} + \dot{M}_{\rm down}\right)
,
\label{mjet}
\\
a(t) &=& a(0) \frac{M_s(0) + M_p}{M_s(0) + M_p - t\left(\dot{M}_{\rm up} + \dot{M}_{\rm down}\right)}
,
\label{ajet}
\\
i(t) &=& i(0) - \frac{3e}{2 \sqrt{1 - e^2}} 
\left(  
\frac{\dot{M}_{\rm down} u_{\rm down} - \dot{M}_{\rm up} u_{\rm up}}{\dot{M}_{\rm down} + \dot{M}_{\rm up}}
\right)
\nonumber
\\
&\times& \sqrt{  \frac{a(0)\left(M_s(0) + M_p\right)}{G M_{p}^2}}
\nonumber
\end{eqnarray}

\begin{equation}
\ \ \ \ 
\times
\ln{ 
\left(
\frac{\left( M_s(0) + M_p \right) \left[M_s(0) - t \left( \dot{M}_{\rm down} + \dot{M}_{\rm up} \right) \right]}
{M_s(0) \left[M_s(0) + M_p - t \left( \dot{M}_{\rm down} + \dot{M}_{\rm up} \right) \right]} 
\right)
}
,
\label{ijet}
\end{equation}

\noindent{whereas} the inclination solution for a test particle secondary is instead

\begin{equation}
i(t) = i(0) - \frac{3e}{2 \sqrt{1 - e^2}}
\sqrt{\frac{a(0)}{G M_s(0)}} 
\left(  
\frac{\dot{M}_{\rm down} u_{\rm down} - \dot{M}_{\rm up} u_{\rm up}}{\dot{M}_{\rm down} + \dot{M}_{\rm up}}
\right)
.
\end{equation}

The solutions show that an eccentric secondary's semimajor axis 
and inclination both {\it evolve monotonically with time}. 
Therefore, the orbital plane always moves towards a pole
unless the orbit is circular.  The higher
the eccentricity, the faster this movement. 
If the jet at the south pole is stronger than that at the north pole 
($\dot{M}_{\rm down} u_{\rm down} > \dot{M}_{\rm up} u_{\rm up}$), then the inclination 
always decreases.  The greater the asymmetry, the faster the inclination changes.  
Also, although the eccentricity remains static, the location of the pericentre 
is a function of time.

\subsubsection{Specific Examples}

Now we consider specific, observationally-motivated examples.  The time-dependent mass-loss rates and velocities of these jets are unknown.  Therefore, let us remove the time dependence of $J$ and $u$ in the following examples.  We now seek to obtain observational constraints, including differences between the north and south jets.

For the stellar mass range of interest here ($M_{\odot} \lesssim 7 M_{\odot}$), we find representative ranges of mass-loss rates of $10^{-10}-10^{-6} M_{\odot}$/yr and ejection velocities of tens to hundreds of km s$^{-1}$.  We obtained these ranges by considering specific examples of observed jets given in table~4 of \cite{podetal2006}, table~3 of \cite{cofetal2008}, table~1 of \cite{meletal2008}, table~1 of \cite{podetal2011}, and table~5 of \cite{elletal2012}\footnote{For more massive stars, typical mass-loss rates often exceed $10^{-5} M_{\odot}$yr$^{-1}$ (see table~2 of \citealt*{cesetal2007}).}.  The observed differences in velocities between both jet components vary but may be approximated by a factor of about two. This factor agrees with the estimate of \cite{namouni2013} and his references.  The asymmetry in the mass-loss rate may similarly be approximated by a factor of a few.

These considerations lead us to select the set
of models in table 2.  The first
9 models feature jets for which the difference
between mass-loss rates and ejecta velocities
for both components is a factor of 2 or 4.  We shrink this factor
to 1.25 and 1.50 for the last 9 models.
For the first 9 models we also vary the mass-loss
rates by 4 orders of magnitude, whereas
for the last 9 models we vary the planet's semimajor 
axis and eccentricity.  In all models $\omega_0 = 0$,
so that equations (\ref{mjet})-(\ref{ijet}) are 
satisfied.  We checked that the output from those equations agree
with output from the full adiabatic set of differential 
equations.  These have further been checked against
the output from the non-adiabatic equations for up 
to $10^4$ orbits of the planet.

We plot the resulting semimajor axis and inclination
variation in Fig. \ref{j1} (for models J1.1-J1.9) and
Fig. \ref{j2} (for models J2.1-J2.9).  These
plots illustrate that if the jets are sustained for
a long enough period of time, then any inclination
may be achieved.  Because $\Delta i \propto u$,
the faster the ejecta velocity, the less mass which
has to be lost in order to achieve the same level of 
orbital excitation.  The semimajor axis evolution is however
independent of $u$ because, in the adiabatic case,
there is no anisotropic contribution to $da/dt$ (further, 
more generally in the
non-adiabatic case, $P_z$ makes no 
contribution to $da/dt$).  Other dependencies apparent
in the plots include $\Delta i \propto e/\sqrt{1-e^2}$ and
$\Delta i \propto \sqrt{a}$.
Fig. \ref{j1} illustrates that for $\dot{M} \approx 10^{-8} M_{\odot}$/yr, 
the jets can produce an inclination variation of about one 
degree in under 1 Myr.  More symmetric jets would hamper this
excitation, but fail to do so in the models of Fig. \ref{j2} because the
mass-loss rate is higher by an order of
magnitude than that of those of Fig.~\ref{j1}.

Finally we caution that, in the more general case with $\omega_0 \ne 0$,
$di/dt$ exhibits more complex behaviour until $\omega$ vanishes.  
Further, in the non-adiabatic case, 
when $a$ is large (typically exceeding hundreds of au), the 
inclination evolution does not behave in such a regular fashion.

\newpage

\section{Conclusion}

We have derived the anisotropic contribution 
(equations \ref{dadtgenP}--\ref{domdtgenP} 
and \ref{dadtP}--\ref{domdtP}) to the orbital 
equations of motion (equations \ref{dadtorig}--\ref{dfdt}) 
for a companion to a primary star 
that is shedding mass in arbitrary directions
with arbitrary velocities.
The relative contribution of the anisotropic terms 
to the overall motion scale as $\sqrt{a}$.  
Because this contribution typically 
vanishes (equations \ref{x0}--\ref{z0}) for a planet in a post-MS 
system anywhere within hundreds
of au of its parent star, we conclude that the isotropic mass
loss approximation is robust for most post-MS planetary studies.
Contrastingly, in nascent planetary systems and in the absence
of other forces, persistent, asymmetric,
sufficiently long-lived bipolar jets may regularly excite 
exoplanet inclinations to a few degrees. 

\section*{Acknowledgements}

This manuscript is dedicated to co-author John D. Hadjidemetriou, whose 
passing came to light the day after submission.  
His distinguished career spanned half of a century, and this paper represents a 
long-planned generalisation of his pioneering 1963 paper results.  He will be 
sorely missed. 

We thank the referee for constructive and insightful comments,
which have improved the context of the manuscript's content 
considerably.  We also thank Lorenzo Iorio for providing us with a copy of 
the paper by Omarov (1962), and Alexander J. Mustill and Cathie J. Clarke for useful 
discussions.  CAT thanks Churchill College for his fellowship.

\onecolumn

\appendix  

\section{}

Here we derive the equations of motion for a jet outflow
of mass in an arbitrary direction, coupled with the background
isotropic mass-loss from the star.  Suppose
the star is losing mass with a realistic flux rate and velocity
given by equations (\ref{Jeq})-(\ref{ueq}) everywhere 
except in a solid angle
bounded by $0 \le \theta_1 \le \theta \le \theta_2 \le \pi$
and $0 \le \phi_1 \le \phi \le \phi_2 \le 2\pi$.  In this region,
the mass-loss rate is constant and uniform and is 
enhanced from $J_0$ by a factor
of $k_J$.  Similarly, the ejecta velocity is constant
and uniform and is enhanced from the polar value
$u_0 \equiv u(\theta = 0)$ by a factor of $k_u$. 
Then

\begin{eqnarray}
P_x &=& \frac{u_0 J_0}{4\pi}
\left[   
\int_{0}^{\phi_1} \cos{\phi} d\phi
\int_{0}^{\theta_1} \left(1 - T \sin^2{\theta}  \right)^{\frac{3}{2}} \sin{\theta} d\theta 
+
k_u k_J  \int_{\phi_1}^{\phi_2} d\phi \int_{\theta_1}^{\theta_2} d\theta
+
\int_{\phi_2}^{2\pi} \cos{\phi} d\phi
\int_{\theta_2}^{\pi} \left(1 - T \sin^2{\theta}  \right)^{\frac{3}{2}} \sin{\theta} d\theta
\right]
\nonumber
\\
&=&
\frac{u_0 J_0}{4\pi}
\left\lbrace
k_u k_J 
\left(\theta_1 - \theta_2\right)  
\left(\phi_1 - \phi_2 \right)
+
\frac{1}{16} \sin{\phi_1}
\left[S_{1-}(\theta_1) + S_{2+}(\theta_1)\right] 
-
\frac{1}{16} \sin{\phi_2}
\left[S_{1+}(\theta_2) - S_{2-}(\theta_2)\right]
\right\rbrace
,
\label{Pxjet}
\\
P_y &=& \frac{u_0 J_0}{4\pi}
\left[   
\int_{0}^{\phi_1} \sin{\phi} d\phi
\int_{0}^{\theta_1} \left(1 - T \sin^2{\theta}  \right)^{\frac{3}{2}} \sin{\theta} d\theta 
+
k_u k_J  \int_{\phi_1}^{\phi_2} d\phi \int_{\theta_1}^{\theta_2} d\theta
+
\int_{\phi_2}^{2\pi} \sin{\phi} d\phi
\int_{\theta_2}^{\pi} \left(1 - T \sin^2{\theta}  \right)^{\frac{3}{2}} \sin{\theta} d\theta
\right]
\nonumber
\\
&=&
\frac{u_0 J_0}{4\pi}
\left\lbrace
k_u k_J 
\left(\theta_1 - \theta_2\right)  
\left(\phi_1 - \phi_2 \right)
+
\frac{1}{16} \left(1 - \cos{\phi_1}\right)
\left[S_{1-}(\theta_1) + S_{2+}(\theta_1)\right] 
-
\frac{1}{16} \left(1 - \cos{\phi_2}\right)
\left[S_{1+}(\theta_2) - S_{2-}(\theta_2)\right]
\right\rbrace
,
\label{Pyjet}
\\
P_z &=& \frac{u_0 J_0}{4\pi}
\left[   
\int_{0}^{\phi_1} d\phi
\int_{0}^{\theta_1} \left(1 - T \sin^2{\theta}  \right)^{\frac{3}{2}} \cos{\theta} d\theta 
+
k_u k_J  \int_{\phi_1}^{\phi_2} d\phi \int_{\theta_1}^{\theta_2} d\theta
+
\int_{\phi_2}^{2\pi} d\phi
\int_{\theta_2}^{\pi} \left(1 - T \sin^2{\theta}  \right)^{\frac{3}{2}} \cos{\theta} d\theta
\right]
\nonumber
\\
&=&
\frac{u_0 J_0}{4\pi}
\left\lbrace
k_u k_J 
\left(\theta_1 - \theta_2\right)  
\left(\phi_1 - \phi_2 \right)
+
\phi_1 S_3(\theta_1)
-
\left(2\pi - \phi_2 \right) S_3(\theta_2)
\right\rbrace
,
\label{Pzjet}
\end{eqnarray}

\noindent{where}, with approximations at small $T$,

\begin{eqnarray}
S_{1\pm}(\theta) &=& 10 - 6T \pm \cos{\theta} \left(5 - 4T + T \cos{2 \theta} \right)\sqrt{4 - 2 T + 2 T \cos{2 \theta} }
\approx
10 \left(1 \pm \cos{\theta}\right) +
\left(-6  \mp \frac{21}{2}\cos{\theta} \pm \frac{9}{2} \cos{\theta}\cos{2\theta}  \right) T
,
\nonumber
\\
S_{2\pm}(\theta) &=& 6 T^{-\frac{1}{2}} \left(T - 1\right)^2  
\ln{ \left[
\frac
{\sqrt{2} \pm\sqrt{2T}}
{\sqrt{2T} \cos{\theta} + \sqrt{2 - T + T\cos{2 \theta}  }}
\right]
}
\approx
6 \left(\pm 1 - \cos{\theta}\right) +
\sin^2{\left(\frac{\theta}{2}\right)} \left(\mp 21  + 2\cos{\theta} \mp \cos{2\theta} \right) T
,
\nonumber
\\
S_{3}(\theta) &=& 3T^{-\frac{1}{2}}\arcsin{\left[\sqrt{T} \sin{\theta} \right]}
+
\sin{\theta} \left(5 - 2 T \sin^2{\theta} \right)
\sqrt{1 - T \sin^2{\theta}  } 
\approx
8 \sin{\theta}  - \left(4 \sin^{3}{\theta} \right)T
.
\nonumber
\end{eqnarray}

\noindent{The} material emanating in the jet generally is constant in neither $k_J$ nor $k_u$.  
In particular, the formation and dissipation of the jet might cause accelerations and 
decelerations which could affect planetary
motion. If all of the stellar mass loss is contained in the jet, then the equations are considerably simplified
and retain a single term each.  In this situation, $P_x = P_y = P_z$.  Similarly, if the vast majority 
of mass loss is in the jet, then the other terms may be neglected.  Additional jets of mass loss may 
be added on to the above relations by splitting the integrals appropriately.  

If there is latitudinal symmetry such that $P_z = 0$ for an ideal jet in the XY plane, then we can attempt to 
solve the adiabatic equations of motion.  Suppose the mass-loss rate of the jet is constant and 
$\dot{M} > 0$, and the mass is lost at a constant speed $u > 0$.  
Then, {\it regardless of the orientation of the jet in the equatorial plane},

\begin{eqnarray}
\frac{da}{dt} &=& -\frac{a(t)}{M_s(t) + M_p} \dot{M}
,
\\
\frac{de}{dt} &=& -\frac{3 \sqrt{1 - e(t)^2}}{2 M_s(t)}
\sqrt{ \frac{a(t)}{G \left(M_s(t) + M_p\right)}} u \dot{M}.
\end{eqnarray}
 
\noindent{These} equations admit complete solutions.  For a finite secondary mass (including a secondary star), the solutions are

\begin{eqnarray}
M_s(t) &=& M_s(0) - t \dot{M}
,
\\
a(t) &=& a(0) \frac{M_s(0) + M_p}{M_s(0) + M_p - t\dot{M}}
,
\label{at1}
\\
e(t) &=& 
\cos{   
\left[
2 {\rm arcsin}\sqrt{\frac{1 - e_0}{2}}
-
\frac{3u}{2} \sqrt{\frac{a_0 \left(M_s(0) +M_p\right)}{G M_{p}^2}}
\ln{ 
\left(  
\frac{M_s(0) \left[M_s(0) + M_p - t \dot{M} \right] }
{ \left( M_s(0) + M_p \right) \left[M_s(0) - t \dot{M} \right]  }
\right)
}
\right]
}
,
\label{et1}
\end{eqnarray}

\noindent{whereas} for a test particle

\begin{equation}
e(t)
=
\cos{   
\left[
- 2 {\rm arcsin}\sqrt{\frac{1 - e_0}{2}}
+
\frac{3ut\dot{M}}{2M_s(0)} \sqrt{\frac{a_0 M_s(0)}{G}}
\frac{1}{M_s(0) - \dot{M}t}
\right]
} 
.
\label{et2}
\end{equation}

Equation (\ref{at1}) shows that a secondary's orbit 
always expands. While doing so, the orbit stretches
and contracts, momentarily becoming circular and flat.
Equations (\ref{et1})-(\ref{et2}) 
can be solved analytically for the times at
which the orbit becomes momentarily flat.  Depending on the
planet's position along its orbit at these times, the planet may
be disrupted or destroyed by the central star.  Let
us consider the first instance in which a
test particle will achieve a flat orbit, and
denote this time as $t_{\rm flat}$.  We find

\begin{equation}
t_{\rm flat}
= \left( \frac{M_s(0)}{\dot{M}} \right)
\left[ 
1 
+
\frac
{ 3u/V_c(0)}
{4 {\rm arcsin} \sqrt{\frac{1 - e_0}{2}} }
  \right]^{-1}
\label{flate}
\end{equation}

\noindent{where} the initial circular velocity
of the particle is $V_c(0) = \sqrt{G M_s(0)/a_0}$.  Equation
(\ref{flate}) demonstrates that $t_{\rm flat}$
is just the time for the star to lose all
its mass, modulo a factor that depends on
$u$, $V_c(0)$ and $e_0$.  In the limit
$u \rightarrow 0$ or $V_c(0) \rightarrow \infty$,
the factor in square brackets becomes
unity, and hence $t_{\rm flat}$ can be
reached only if the entire star is dissipated.
Alternatively, in the limit $u \rightarrow \infty$,
the orbit becomes flat immediately.  The limit
$V_c(0) \rightarrow 0$ does not apply
because then the particle would reside in the 
nonadiabatic regime.  Initially circular
orbits ($e_0 = 0$) take the longest to
achieve a flat orbit but can still easily 
reach that orbit.  Additionally, the ratio in 
parenthesis is proportional to the reciprocal of
the mass loss adiabaticity index from \cite{veretal2011},
and this index could be used to help ensure
adiabaticity is maintained (so that equation 
\ref{flate} remains valid) as the eccentricity
increases.

While
stretching and contracting, the pericentre of the orbit
evolves according to

\begin{equation}
\frac{d\omega}{dt}
=
\frac{3 \sqrt{1-e(t)^2}}{2 e(t) M_s(t)}
\sqrt{ \frac{a(t)}{G \left(M_s(t) + M_p\right)}} u \dot{M}
.
\end{equation}

\label{lastpage}

\end{document}